\RequirePackage{fix-cm}
\documentclass[smallcondensed]{svjour3}     
\smartqed

\usepackage{graphicx}
\usepackage{cite}
\usepackage{hyperref}
\usepackage{amsmath}
\usepackage{float}
\usepackage{upgreek}
\usepackage{amssymb}
\usepackage{mathtools}
\usepackage[linesnumbered,ruled,vlined]{algorithm2e}
\SetKwInput{KwInput}{Input}                
\SetKwInput{KwOutput}{Output}              
\usepackage{booktabs}
\usepackage{tabu} 
\usepackage{dcolumn}
\newcolumntype{A}{D{.}{.}{2.3}}
\usepackage[title]{appendix}
\usepackage{placeins}
\usepackage{subcaption}

\journalname{Multibody System Dynamics}

\def\vec#1{\ensuremath{\mathchoice
                     {\mbox{\boldmath$\displaystyle\mathbf{#1}$}}
                     {\mbox{\boldmath$\textstyle\mathbf{#1}$}}
                     {\mbox{\boldmath$\scriptstyle\mathbf{#1}$}}
                     {\mbox{\boldmath$\scriptscriptstyle\mathbf{#1}$}}}}
                     
\begin{document}

\title{A low order, torsion-deformable spatial beam element based on the absolute nodal coordinate formulation and Bishop frame\thanks{This is a preprint of an article published in \emph{Multibody System Dynamics}. The final authenticated version is available online at: \href{https://doi.org/10.1007/s11044-020-09765-7}{https://doi.org/10.1007/s11044-020-09765-7}.}}

\author{Mehran Ebrahimi \and Adrian Butscher \and Hyunmin Cheong}

\institute{M. Ebrahimi 
			\at  Autodesk Research, 661 University Avenue, Toronto, ON M5G 1M1, Canada  \\
              \email{mehran.ebrahimi@autodesk.com}           
           \and
           A. Butscher 
           \at Autodesk Research, 661 University Avenue, Toronto, ON M5G 1M1, Canada  \\
           \email{adrian.butscher@autodesk.com}
           \and
           H. Cheong 
           \at Autodesk Research, 661 University Avenue, Toronto, ON M5G 1M1, Canada  \\
           \email{hyunmin.cheong@autodesk.com}
}

\date{Received: date / Accepted: date}

\maketitle

\begin{abstract}
Heretofore, The Serret-Frenet frame has been the ubiquitous choice for analyzing the elastic deformations of beam elements. It is well-known that this frame is undefined at the inflection points and straight segments of the beam where its curvature is zero, leading to singularities and errors in their numerical analysis. On the other hand, there exists a lesser-known frame called Bishop which does not have the caveats of the Serret-Frenet frame and is well-defined everywhere along the beam center-line. Leveraging the Bishop frame, in this paper, we propose a new spatial, singularity-free low order beam element based on the absolute nodal coordinate formulation for both small and large deformation applications. This element, named ANCF14, has a constant mass matrix and can capture longitudinal, transverse (bending) and torsional deformations. It is a two-noded element with 7 degrees of freedom per node, which are global nodal coordinates, nodal slopes and their cross-sectional rotation about the center-line. The newly developed element is tested through four complex benchmarks. Comparing the ANCF14 results with theoretical and numerical results provided in other studies confirms the efficiency and accuracy of the proposed element.
\end{abstract}

\section{Introduction}
\label{sec:introduction}
Multibody systems (MBSs) are mechanical assemblies consisting of interconnected rigid and flexible components that may undergo large rotations and displacements, as well as large deformations in their flexible parts. Simulation of MBSs has been an attractive subject of interest in engineering literature and numerous studies have been devoted to improving the accuracy and efficiency of the numerical techniques for simulating the behavior of such systems \cite{shabana2020dynamics,geradin2001flexible, schiehlen1997multibody, shabana1997flexible}. This paper presents the development of a new singularity-free three-dimensional beam element capable of handling longitudinal, bending and torsional deformations with a reduced number of degrees of freedom (DOF), compared to the original spatial elements introduced in \cite{shabana2001} and \cite{yakoub2001three}, based on the \emph{absolute nodal coordinate formulation} (ANCF) and the \emph{Bishop} frame. 

Several formulations have been proposed for analyzing the nonlinear deformation of beams in static and dynamic MBSs. The \emph{floating frame of reference} \cite{canavin1977floating}, the \emph{incremental finite element} \cite{rankin1986element, shabana1996finite}, the \emph{large rotation vector} \cite{simo1985finite}, the \emph{geometrically exact beam} \cite{bauchau1987large, simo1985finite, simo1986three, simo1986dynamics, geradin2001flexible} and the \emph{absolute nodal coordinate formulation} \cite{Shabana1996, shabana2001, yakoub2001three, Qiang2010, Gerstmayr2013} are the most widely used. In the floating frame of reference method, two coordinate systems are associated with each flexible body; one capturing its rigid-body motion (i.e. large rotations and displacements) and the other for its elastic deformations with respect to the first frame. This approach leads to zero strains for a pure rigid body motion and yields the same stiffness matrix as that in the linear elasticity finite element method. However, it is limited to small deformations and results in a nonlinear time-dependent mass matrix, centrifugal forces and Coriolis forces in the dynamic equations.

The incremental finite element formulation, on the other hand, can model large elastic deformations. In this technique, large rotations of elements are described incrementally by a sequence of infinitesimal rotations and elements are configured using their nodal coordinates and angles. This method leads to non-zero strains in the case of rigid-body rotations which makes it inapplicable to the problems with geometric nonlinearities. The large rotation vector approach can handle both large deformations and rotations by representing the configuration of each element using their global nodal coordinates and rotations. This method, however, leads to a time-dependent mass matrix and an excessive shear force in the element's cross-section \cite{shabana1998computer}.  The geometrically exact beam formulation is also able to account for large deformations in beams, but similar to all other aforementioned methods results in a non-constant mass matrix that needs to be updated at each time-step of the solution process \cite{gerstmayr2008correct}.

The ANCF describes the configuration of finite elements by their nodal \emph{positions} and \emph{slopes} in the global coordinate system. The ANCF has been extensively used for simulating beams, plates \cite{dmitrochenko2003generalization, mikkola2006development} and solids \cite{olshevskiy2014three}, in static and dynamic scenarios and is capable of describing rigid-body modes and solving large deformation problems. Unlike previous methods, the ANCF results in a constant mass matrix for the elements and eliminates the nonlinear terms of centrifugal and Coriolis forces in the equations of motion. These can alleviate the computational costs associated with solving the dynamic equations and significantly simplify the equations involved in the sensitivity analysis for optimizing MBSs \cite{ebrahimi201982}. Also, constraint equations defining the coupling between the bodies can be expressed in simple terms, thus facilitating the development of different joint types. 

The original ANCF was proposed by Shabana for two-dimentional beam elements in 1996 \cite{Shabana1996}. Since then, it has gone through many improvements regarding its accuracy and efficiency. A thorough review of the various ANCF-based beam and plate elements can be found in \cite{Gerstmayr2013}. A subtle and complex deformation mode in beam elements is the torsion of their cross-section about their center-line and many techniques have been developed to analyze it using the ANCF.  Von Dombrowski \cite{von2002analysis} introduced torsional effects by including parameterized rotation into the equations of motion. This, however, leads to a time-dependent mass matrix for the elements, which compromises the aforementioned benefits of having a constant mass matrix in the ANCF. In Yakoub and Shabana\cite{yakoub2001three}, Dmitrochenko and Pogorelov \cite{dmitrochenko2003generalization} and Yoo et al. \cite{yoo2005new}, different sets of DOF are used to represent the configuration of three-dimensional shear and torsion deformable beam elements. These proposals, like many other ANCF-based formulations, rely on the \emph{Serret-Frenet} (SF) frame \cite{burke1985applied} to define the transverse bending deformation and cross-sectional torsion (twist). 

A notorious characteristic of the SF frame for spatial curves is that it is undefined at \emph{inflection points} (points with zero curvature) and straight segments of the curve \cite{hanson1995parallel}. Furthermore, while sweeping through the inflection points, the SF frame exhibits unnecessary rotation about the curve's tangent vector that leads to an excessive torsional energy and a sudden flip of this frame. Therefore, singularities may arise in solving the governing equations. 

To cope with these issues, we propose a new beam element based on the ANCF that uses the Bishop frame \cite{bishop1975there} for describing the kinematics of the beam's cross-section. The Bishop frame has a sound mathematical foundation in terms of the center-line's equation and is well-defined even in the case of a vanishing curvature, thus delivering a singularity-free formulation for characterizing the beam configuration. This beam element can handle torsional deformations and requires fewer number of DOF compared to its ANCF-based counterparts, which could significantly reduce the relative computational costs associated with solving the dynamic equations. Similar to the element proposed by Yoo et al. in \cite{yoo2005new}, global nodal coordinates and their slopes are utilized to define the center-line and the cross-sectional torsion is determined by a rotation angle about the beam center-line. This beam element follows the Euler-Bernoulli beam theory and has 7 DOF per node, thus 14 for its two-noded version, and is named ANCF14.

The remainder of the paper is organized as follows. In the next section, the SF and Bishop frames for spatial curves and their differences are explained.  Using the definition of the Bishop frame in Section \ref{sec:frame}, the kinematics of our newly proposed element are provided in Section \ref{sec:kinematics}. Section \ref{sec:motion} introduces the dynamic equations of ANCF14 and formulates its kinetic and potential energies, as well as its mass matrix. Section \ref{sec:motion} also derives the equations of the elastic potential energy and its virtual work for ANCF14. Four numerical examples are presented in Section \ref{sec:examples}, followed by concluding remarks in Section \ref{sec:conclusion}.

\section{Coordinate systems for spatial curves}
\label{sec:frame}

In the ANCF theory, the properties of the beam center-line fully-determine the longitudinal deformation of the beam. For the bending deformation and cross-sectional torsion, on the other hand, a local coordinate system and its evolution between the two ends of the beam are required. Therefore, it is of substantial importance to be able to consistently define a local coordinate system along the beam and track its evolution to properly model the beam's elastic deformations. 

For a spatially parameterized curve such as $\vec{r}(x)$, potentially infinitely many \emph{adapted} Cartesian coordinate systems can be defined. If their \emph{tangent}, \emph{normal} and \emph{bi-normal} axes are represented respectively by $\vec{t}(x)$, $\vec{n}(x)$ and $\vec{b}(x)$, considering their orthonormality, there exist scalar functions $\kappa_1(x)$, $\kappa_2(x)$ and $\tau(x)$ such that \cite{audoly2000elasticity}  

\begin{equation}\label{eq:systemdiff}
\begin{aligned}
\vec{t}'(x) &= \kappa_1(x) \vec{n} (x) + \kappa_2(x) \vec{b}(x) \\
\vec{n}'(x) &= -\kappa_1(x) \vec{t}(x) + \tau(x) \vec{b}(x) \\
\vec{b}'(x) &= -\kappa_2(x) \vec{t}(x) - \tau(x) \vec{n} (x)
\end{aligned} 
\end{equation}

\noindent where \emph{prime} denotes differentiation with respect to $x$ and

\begin{equation}
\kappa_1(x) = \vec{t}' (x) \cdot \vec{n}(x), \: \kappa_2(x) = \vec{t}' (x) \cdot \vec{b}(x), \: \tau(x) = \vec{n}' (x) \cdot \vec{b}(x)
\end{equation}

\noindent In a compact form, Eq. (\ref{eq:systemdiff}) can be written as

\begin{equation}\label{eq:diffeqgen}
\begin{aligned}
\vec{t}'(x) & = \vec{\Omega}(x) \times \vec{t}(x) \\
\vec{n}'(x) & = \vec{\Omega}(x) \times \vec{n}(x) \\
\vec{b}'(x) & = \vec{\Omega}(x) \times \vec{b}(x)
\end{aligned} 
\end{equation}

\noindent in which $\vec{\Omega}(x)$ is the so-called \emph{Darboux} vector of the frame, given by

\begin{equation}\label{eq:darbouxgen}
\vec{\Omega}(x) = \tau(x) \vec{t}(x) - \kappa_2(x) \vec{n}(x) + \kappa_1(x) \vec{b}(x)
\end{equation}

\begin{figure*}
\centering
\includegraphics[width=0.8\textwidth]{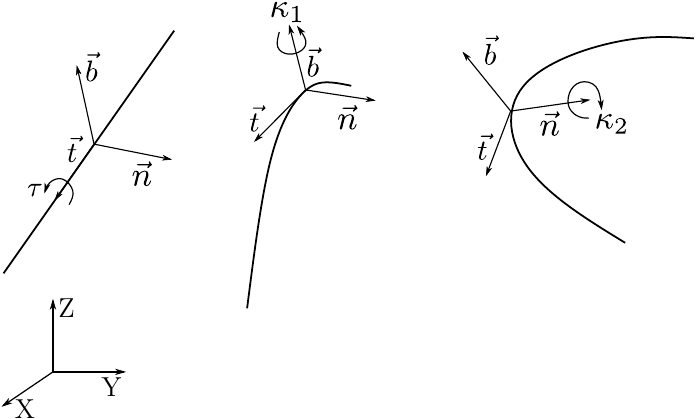}
\caption{The meaning of $\kappa_1$, $\kappa_2$ and $\tau$ for a frame}
\label{fig:generalframe}
\end{figure*}

As depicted in Figure \ref{fig:generalframe}, $\kappa_1 (x)$, $\kappa_2(x)$ and $\tau (x)$ describe the rates of rotation of the frame about $\vec{b}(x)$, $\vec{n}(x)$ and $\vec{t}(x)$, respectively. In fact, $\kappa_1(x)$ and $\kappa_2(x)$ represent the curve's curvatures and $\tau(x)$ indicates the frame's \emph{twist} (torsion) about the tangent vector. 

Eq. (\ref{eq:diffeqgen}) is a system of differential equations describing the evolution of a frame along the curve and depending on the conditions (i.e. constraints) imposed on it, different coordinate systems can be generated for the curve. By construction, the SF and Bishop frames are defined by assuming $\kappa_2(x)=0$ and $\tau(x)=0$, respectively. Heretofore, in the ANCF-based beam models and perhaps most of the other beam theories, the SF frame has been the standard choice, since it can be easily computed analytically through Eq. (\ref{eq:systemdiff}), using solely the curve equation. For a three-dimensional parameterized curve (e.g. beam center-line), the SF frame behaves unpredictably at inflection points where the second derivative of the curve is zero \cite{hanson1995parallel}. The Bishop frame, however, is well-defined even for the straight segments of the curve. In this section, these two frames are briefly introduced.

\subsection{Serret-Frenet frame}
\label{subsec:frenet}

Setting $\kappa_2(x)=0$ reduces Eq. (\ref{eq:systemdiff}) to

\begin{equation}\label{eq:SFsystemdiff}
\begin{aligned}
\vec{t}'(x) &= \kappa(x) \vec{n}(x) \\
\vec{n}'(x) &= -\kappa(x) \vec{t}(x) + \tau(x) \vec{b}(x) \\
\vec{b}'(x) &= -\tau(x) \vec{n}(x)
\end{aligned} 
\end{equation}

\noindent with $\kappa(x) \coloneqq \kappa_1(x)$. The Darboux vector of the SF frame also simplifies to 

\begin{equation}\label{eq:frenetdarboux}
\vec{\Omega}_{\mathrm{SF}}(x) = \kappa(x) \vec{b}(x) + \tau(x) \vec{t}(x)
\end{equation}

\noindent For the SF frame, $\kappa (x)$ and $\tau (x)$ are analytically defined by

\begin{equation}\label{eq:kappaandtau}
\begin{aligned}
\kappa (x) &= \frac{\left\| \vec{r}'(x) \times \vec{r}''(x) \right\|}{\left\| \vec{r}'(x) \right\|^3} \\
\tau (x) &= \frac{\left( \vec{r}'(x) \times \vec{r}''(x) \right)\cdot \vec{r}'''(x)}{\left\| \vec{r}'(x) \times \vec{r}''(x) \right\|^2}
\end{aligned} 
\end{equation} 

\noindent The three axes $\vec{t}(x)$, $\vec{n}(x)$ and $\vec{b}(x)$ can be computed using Eq. (\ref{eq:SFsystemdiff}) through Eq. (\ref{eq:kappaandtau}) as

\begin{equation}\label{eq:frenetvectors}
\begin{aligned}
\vec{t}(x) &= \frac{\vec{r}'(x)}{\left\| \vec{r}'(x) \right\|} \\
\vec{b}(x) &= \frac{\vec{r}'(x) \times \vec{r}''(x)}{\left\| \vec{r}'(x) \times \vec{r}''(x) \right\|} \\
\vec{n}(x) &= \vec{b}(x) \times \vec{t}(x)
\end{aligned} 
\end{equation}

\begin{figure*}[ht]
\centering
\includegraphics[width=0.9\textwidth]{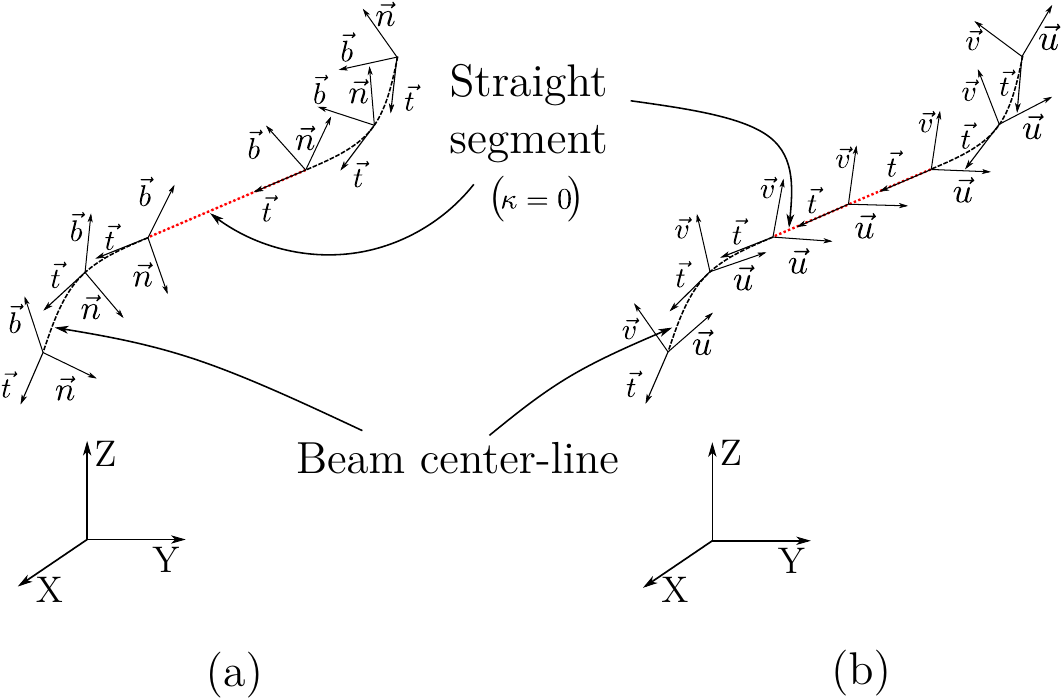}
\caption{The evolution of (a) SF frame and (b) Bishop frame along a spatial curve}
\label{fig:frenetvsbishop}
\end{figure*}

The normal and bi-normal vectors define the normal plane to the center-line. For Euler-Bernoulli beams, this plane coincides with the cross-section. It is clear from Eq. (\ref{eq:frenetvectors}) that the bi-normal vector and, subsequently, the normal vector are undefined where $\vec{r}  '' (x) = \vec{0}$, thus causing numerical problems when used for analyzing the beam deformations. Figure \ref{fig:frenetvsbishop}(a) illustrates the evolution of the SF frame for a spatial curve. Notice that the frame is undefined on the straight segment of the curve and swings abruptly while passing through this zero-curvature region.

\subsection{Bishop frame}
\label{subsec:bishop}

The Darboux vector of the SF frame, Eq. (\ref{eq:frenetdarboux}), contains a component along the tangential direction, which means that a coordinate system would twist if it is transported along a curve according to this vector.  The Bishop frame, on the other hand, provides a way to move the frame with a uniform zero twist along the same curve. Hence, it leads to the most geometrically natural evolution of the local frame \cite{bergou2008discrete}. To distinguish between this frame and the SF frame, let $\vec{t}(x)$, $\vec{u}(x)$ and $\vec{v}(x)$ correspond to the three axes of the frame at $x$. By definition, in the Bishop frame $\tau(x) =  0 $ \cite{bishop1975there}, resulting the following differential equations for this coordinate system.

\begin{equation}\label{eq:RMsystemdiff}
\begin{aligned}
\vec{t}'(x) &= \kappa_1(x) \vec{u}(x) + \kappa_2(x) \vec{v}(x) \\
\vec{u}'(x) &= -\kappa_1(x) \vec{t}(x) \\
\vec{v}'(x) &= -\kappa_2(x) \vec{t}(x)
\end{aligned} 
\end{equation}

\noindent Eq. (\ref{eq:RMsystemdiff}) can be further shortened to

\begin{equation}\label{eq:bishopdiffeq}
\begin{aligned}
\vec{t}'(x) &= \vec{\Omega}_{\mathrm{Bishop}}(x) \times \vec{t}(x) \\
\vec{u}'(x) &= \vec{\Omega}_{\mathrm{Bishop}}(x) \times \vec{u}(x) \\
\vec{v}'(x) &= \vec{\Omega}_{\mathrm{Bishop}}(x) \times \vec{v}(x)
\end{aligned} 
\end{equation}

\noindent The Bishop frame's Darboux vector is thus expressed as 

\begin{equation}\label{eq:bishopdarboux}
\begin{aligned}
\vec{\Omega}_{\mathrm{Bishop}}(x) = \kappa_1(x)\vec{v}(x)-\kappa_2(x)\vec{u}(x)
\end{aligned}
\end{equation}

This vector has no tangential component and moves the frame along the curve with no twist, as demonstrated in Figure \ref{fig:frenetvsbishop}(b). The vector $\vec{t}(x)$ of the Bishop and SF frames are the same and can be computed analytically utilizing Eq. (\ref{eq:frenetvectors}). To find $\vec{u}(x)$ and $\vec{v}(x)$, Eq. (\ref{eq:RMsystemdiff}) needs to be solved. Doing so is not as straightforward as that for the SF frame, so a number of approximation techniques, such as the \emph{rotation method}, the \emph{double reflection method} and the \emph{numerical integration method} have been developed \cite{wang2008computation}, the first one being the most widely used. In this paper, we adopt the rotation method proposed by Hanson and Ma \cite{hanson1995parallel}.

\subsubsection{Finding the Bishop frame along a spatial curve}
\label{subsubsec:findingbishop}

Eq. (\ref{eq:RMsystemdiff}) is an initial-value problem and given an initial condition at $x=0$, its solution can be uniquely found along the curve. Suppose at $x=0$, a Cartesian coordinate system denoted by $\left\{ \vec{t}^{0}, \vec{u}^{0}, \vec{v}^{0} \right\}$ is defined and the goal is to find the Bishop frame at the desired $N$ locations along the curve. This is an admissible assumption, as for solving the dynamic equations of beams, the physical quantities of interest (explained in the forthcoming sections) are integrated numerically along the center-line. Hence, having the Bishop frame only at the integration points (e.g. Gaussian quadrature points) is adequate for the sake of this study. The algorithm proceeds as described in Algorithm \ref{alg:rotationBP}.

\vspace{1em}
\begin{algorithm}[H]
\caption{The Bishop frame for a selected number of points along a spatial curve}
\label{alg:rotationBP}
\SetAlgoLined
\KwInput{Curve equation $\vec{r}(x)$, $\left\{ \vec{t}^{0}, \vec{u}^{0}, \vec{v}^{0} \right\}$ and the desired $x_i$ locations}
\KwOutput{Bishop frame $\left\{ \vec{t}^{i}, \vec{u}^{i}, \vec{v}^{i} \right\}$ for $i=1,\cdots,N$}
 \For{$i=1, \cdots, N$}{
  $\vec{t}^{i} \leftarrow \vec{r}^{'}(x_i) / \| \vec{r}^{'}(x_i)\|$\;
  $\vec{n} \leftarrow \vec{t}^{i-1} \times \vec{t}^{i}$\;
  \eIf{$\| \vec{n} \| = 0$}{
   $\vec{u}^{i} \leftarrow \vec{u}^{i-1}$\;
   $\vec{v}^{i} \leftarrow \vec{v}^{i-1}$\;
   }{
   $\vec{n} \leftarrow \vec{n} / \| \vec{n} \|$\;
   $\phi \leftarrow \arccos \left(\vec{t}^{i-1} \cdot \vec{t}^{i} \right)$\;
   $\vec{u}^i \leftarrow \vec{u}^{i-1} \cos(\phi) + \left(\vec{n} \times \vec{u}^{i-1} \right) \sin(\phi) + \vec{n} \left( \vec{n} \cdot \vec{u}^{i-1} \right) \left( 1- \cos(\phi) \right) $\;
   $\vec{v}^i \leftarrow \vec{v}^{i-1} \cos(\phi) + \left(\vec{n} \times \vec{v}^{i-1} \right) \sin(\phi) + \vec{n} \left(\vec{n} \cdot \vec{v}^{i-1} \right) \left( 1- \cos(\phi) \right) $\; 
  }
 }
\end{algorithm}
\vspace{1em}

\begin{figure*}[ht]
\centering
	\begin{subfigure}{0.49\textwidth}
        \includegraphics[width=\textwidth]{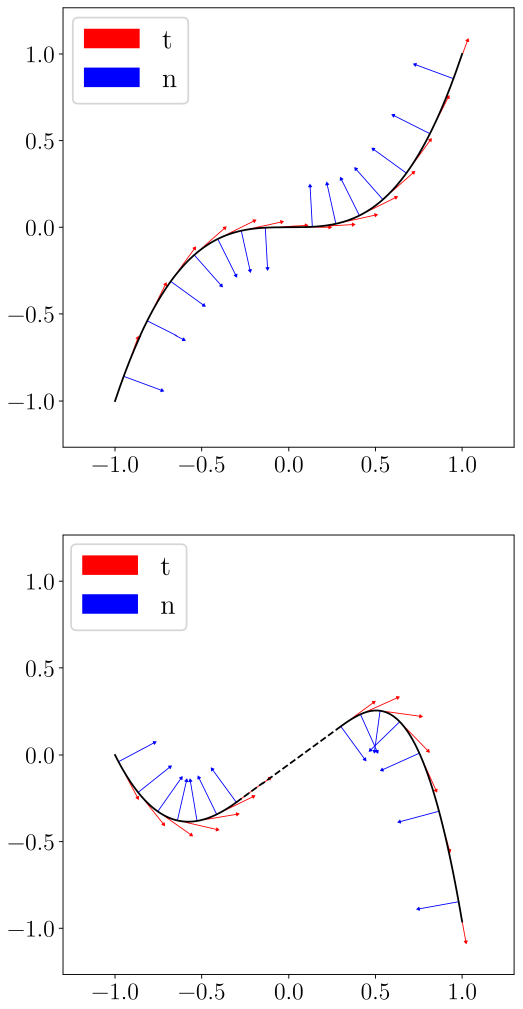}
        \caption{}
        \label{subfig:serret}
    \end{subfigure}
    \begin{subfigure}{0.49\textwidth}
        \includegraphics[width=\textwidth]{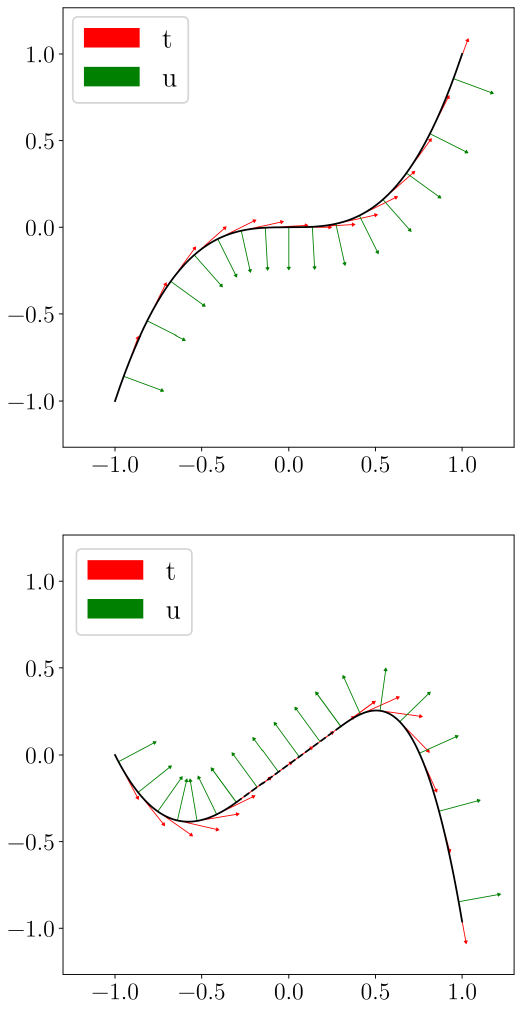}
        \caption{}
        \label{subfig:bishop}
    \end{subfigure}
    \caption{(a) SF and (b) Bishop frames for two planar curves with an inflation point (top) and a straight line segment (bottom)}
    \label{fig:SFvsBP} 
\end{figure*}

In fact, the Bishop frame is obtained through the successive alignment of the tangent vectors along the beam. Of course, increasing the number of intermediate points would improve the accuracy of the approximation. Figure \ref{fig:SFvsBP} depicts the SF and Bishop frames (employing respectively Eq. (\ref{eq:frenetvectors}) and Algorithm \ref{alg:rotationBP}) for two piecewise third-order polynomial planar curves. 

\FloatBarrier
\section{Kinematics of ANCF14}
\label{sec:kinematics}

\begin{figure*}
\centering
\includegraphics[width=0.75\textwidth]{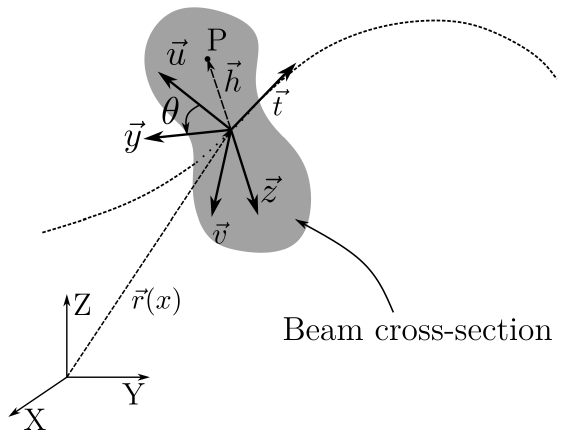}
\caption{The material and Bishop frames for a cross-section}
\label{fig:crosssection}
\end{figure*}

To account for the cross-sectional torsion, a consistent way of tracking the rotation of the cross-section along the beam center-line is required. This can be achieved by leveraging the Bishop frame. Depicted in Figure \ref{fig:crosssection}, assume that at each $x$, a coordinate system called the \emph{material} frame is rigidly attached to the cross-section that rotates along with it about the center-line. Suppose $\theta (x)$ denotes this rotation relative to the Bishop frame at $x$. The material frame consists of an axis coinciding with $\vec{t}(x)$ and two orthonormal vectors $\vec{y}(x)$ and $\vec{z}(x)$ perpendicular to $\vec{t}(x)$. Having the Bishop frame at $x$, $\vec{y}(x)$ and $\vec{z}(x)$ can be computed by

\begin{equation}\label{eq:materialfrombishop}
\begin{aligned}
\vec{y}(x) & = \cos \left( \theta (x) \right) \vec{u}(x) + \sin \left( \theta (x) \right) \vec{v}(x)\\
\vec{z}(x) & = -\sin \left( \theta (x) \right) \vec{u}(x) + \cos \left( \theta (x) \right) \vec{v}(x)
\end{aligned} 
\end{equation}

Similar to the SF and Bishop frames, the equations for evolving the material frame along the center-line can be written as

\begin{equation}\label{eq:materialdiffpre}
\begin{aligned}
\vec{t}'(x) &= \vec{\Omega}_{\mathrm{m}}(x) \times \vec{t}(x) \\
\vec{y}'(x) &= \vec{\Omega}_{\mathrm{m}}(x) \times \vec{y}(x)\\
\vec{z}'(x) &= \vec{\Omega}_{\mathrm{m}}(x) \times \vec{z}(x)
\end{aligned} 
\end{equation}

\noindent with $\vec{\Omega}_{\mathrm{m}}(x)$ as the Darboux vector of this coordinate system 

\begin{equation}\label{eq:materialdarboux}
\begin{aligned}
\vec{\Omega}_{\mathrm{m}} (x) & = \left(\vec{y}'(x) \cdot \vec{z}(x)\right) \vec{t}(x) - \left(\vec{t}'(x) \cdot \vec{z}(x)\right) \vec{y}(x) + \left(\vec{t}'(x) \cdot \vec{y}\right) \vec{z}(x) \\
&= \tau_m(x) \vec{t}(x) -\gamma_2(x) \vec{y}(x) + \gamma_1(x) \vec{z}(x)
\end{aligned}
\end{equation}

\noindent where $\gamma_1$ and $\gamma_2$ are the material frame's curvatures and $\tau_m$ is its torsion about $\vec{t}$. Following Eq. (\ref{eq:materialfrombishop}) and Eq. (\ref{eq:materialdarboux}), $\tau_{\mathrm{m}}(x)$ reads

\begin{equation}
\begin{aligned}
\tau_{\mathrm{m}}(x) &= \vec{y} ' (x) \cdot \vec{z}(x) \\
&= \theta  ' (x)  - 2 \: \theta' (x) \sin(\theta(x)) \cos(\theta(x)) \left( \vec{u}(x) \cdot \vec{v}(x)  \right) \\ 
&- \sin^2(\theta(x)) \left(\vec{u}(x) \cdot \vec{v} ' (x) \right) + \cos^2(\theta(x)) \left(\vec{v}(x) \cdot \vec{u} ' (x) \right)
\end{aligned} 
\end{equation}


\noindent Considering the orthonormality of vectors $\vec{u}(x)$ and $\vec{v}(x)$, and the fundamental characteristic of the Bishop frame that is $\vec{u}(x) \cdot \vec{v} '(x) = \vec{v}(x) \cdot \vec{u} '(x) = 0$, the final expression for $\tau_{\mathrm{m}}(x)$ becomes

\begin{equation}\label{eq:torsiondef}
\tau_{\mathrm{m}}(x) = \theta'(x)
\end{equation}

Accordingly, $\theta (x)$ describes both the cross-sectional rotation (mechanical torsion) and the coordinate system's twist (geometrical torsion) and is taken as a nodal degree of freedom for ANCF14, leading to the following \emph{time-dependent} vector of nodal coordinates for this element

\begin{equation}\label{eq:nodalcoords}
\begin{aligned}
\vec{q}(t) & \coloneqq \begin{bmatrix}
\vec{r}_i^\mathrm{T} \quad & \quad \partial\vec{r}_i^\mathrm{T}/\partial x \quad & \quad \theta_i \quad & \quad \vec{r}_j^\mathrm{T} \quad & \quad \partial\vec{r}_j^\mathrm{T} / \partial x \quad & \quad \theta_j
\end{bmatrix}^\mathrm{T}
\end{aligned}
\end{equation}

\noindent where $i$ and $j$ are the two nodes of the element, and $\vec{r}$, $\partial\vec{r} / \partial x$ and $ \theta$ represent respectively the global coordinates, global slopes and cross-sectional rotation about the center-line at those nodes. An ANCF14 beam element is illustrated in Figure \ref{fig:ANCF14}. 

Following Figure \ref{fig:crosssection}, if the position of a point P on the cross-section in the material frame is denoted by $\vec{h} \coloneqq \begin{bmatrix} 0 \quad & \quad \bar{y} \quad & \quad \bar{z}\end{bmatrix}^{\mathrm{T}}$, its global position $\vec{r}_{\mathrm{P}}(x)$ can be calculated through

\begin{equation}\label{eq:position}
\begin{gathered}
\vec{r}_{\mathrm{P}}(x,t)= \vec{r}(x,t)+ \mathbf{R}(x,t)\vec{h} =  \mathbf{S}(x)\vec{q}(t) + \mathbf{R}(x,t)\vec{h} \\
\end{gathered}
\end{equation}

\noindent in which $\mathbf{R} \coloneqq \begin{bmatrix} \vec{t} \quad & \quad \vec{y} \quad & \quad \vec{z} \end{bmatrix}$ represents the material frame and $\mathbf{S}$ is the \emph{time-independent} shape function matrix stated as

\begin{equation}\label{eq:shapefunction}
\begin{gathered}
\mathbf{S}(x) = \begin{bmatrix}
s_1 \mathbf{I}_{3} \quad & \quad s_2 \mathbf{I}_{3} \quad & \quad \vec{0}_3 \quad & \quad s_4 \mathbf{I}_{3} \quad & \quad s_5 \mathbf{I}_{3} \quad & \quad \vec{0}_3
\end{bmatrix}\\
s_1 = 1-3(\frac{x}{l})^2+2(\frac{x}{l})^3, \quad s_2 = l\left((\frac{x}{l})-2(\frac{x}{l})^2+(\frac{x}{l})^3\right) \\
s_3 = 3(\frac{x}{l})^2-2(\frac{x}{l})^3, \quad s_4 = l\left((\frac{x}{l})^3-(\frac{x}{l})^2\right)
\end{gathered}
\end{equation}

\noindent with $\vec{0}_3 \coloneqq \begin{bmatrix} 0 \quad & \quad 0 \quad & \quad 0\end{bmatrix}^{\mathrm{T}}$. In Eq. (\ref{eq:shapefunction}), $l$ is the element’s length in the undeformed configuration and $x$ is measured from Node $i$ in the undeformed configuration. The rotation angle $\theta(x)$ is interpolated linearly from Node $i$ to Node $j$

\begin{equation}\label{eq:rotationinterpolation}
\begin{gathered}
\theta(x, t) = \left(1-\frac{x}{l} \right)\theta_i + \frac{x}{l} \theta_j = \mathbf{\bar{S}}(x)\vec{q}(t) \\
\end{gathered}
\end{equation}

\noindent leading to the following \emph{time-independent} shape function matrix for the cross-sectional rotation

\begin{equation}\label{eq:rotationshapefunc}
\begin{gathered}
\mathbf{\bar{S}}(x) = \begin{bmatrix} 0 \quad & \quad 0 \quad & \quad 0 \quad & \quad 0 \quad & \quad 0 \quad & \quad 0 \quad & \quad 1-x/l \quad & \quad 0 \quad & \quad 0 \quad & \quad 0 \quad & \quad 0 \quad & \quad 0 \quad & \quad 0 \quad & \quad x/l \end{bmatrix}
\end{gathered}
\end{equation}

\begin{figure*}
\centering
\includegraphics[width=1.0\textwidth]{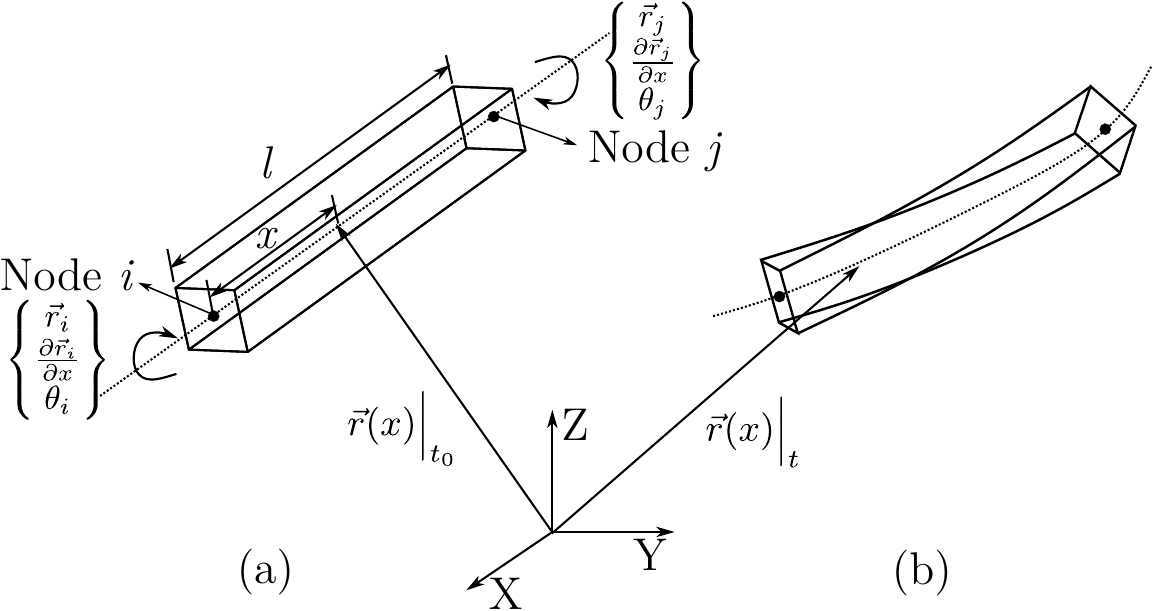}
\caption{The ANCF14 beam element in the (a) undeformed and (b) deformed configurations}
\label{fig:ANCF14}
\end{figure*}

\section{Multibody dynamics governing equations}
\label{sec:motion}

Due to \emph{Hamilton's principle}, the motion of a dynamic system can be determined by finding the stationary values of its \emph{action integral}. For a \emph{conservative} constrained MBS the continuous action integral for time $t \in \left[0,T \right] $ is given by
\begin{equation}\label{eq:action}
S (\vec{q}) = \int_0^T \left( L \left( \vec{q}, \dot{\vec{q}} \right) -
\vec{\lambda}\cdot \vec{g} (\vec{q})
 \right) dt
\end{equation}

\noindent where $L( \vec{q},\dot{\vec{q}})$ is the MBS's Lagrangian and $\vec{g} (\vec{q}) \in \mathbb{R}^m$ represents $m$ \emph{holonomic} constraint (joints) equations between different bodies in the assembly. In this equation, $\vec{q} \in \mathbb{R}^n$ and $\dot{\vec{q}}\in \mathbb{R}^n$ are the vectors of all DOF of the system and their time derivatives, respectively. Also, $\vec{\lambda} \in \mathbb{R}^m $ describes the vector of Lagrange multipliers associated with the $m$ constraints. The Lagrangian function $L$ is the difference between the system's kinetic energy $T$ and potential energy $U$
\begin{equation}\label{eq:lagrangian}
L \left( \vec{q}, \dot{\vec{q}} \right) = T(\dot{\vec{q}})-U(\vec{q})
\end{equation}

Taking the variation of Eq. (\ref{eq:action}) with respect to $\vec{q}$, setting it to zero and incorporating the constraint equations lead to the \emph{Euler-Lagrange} equations governing the motion of MBSs in space and time

\begin{equation}\label{eq:govern}
\begin{aligned}
\frac{d}{dt}\left( \frac{\partial T}{\partial \dot{\vec{q}}}\right) - \frac{\partial T}{\partial \vec{q}} + \frac{\partial U}{\partial \vec{q}} + \frac{\partial \vec{g}}{\partial \vec{q}}^{\mathrm{T}} \vec{\lambda}  &=\vec{0} \\
\vec{g}&=\vec{0}
\end{aligned}
\end{equation}

In this equation, $\partial U / \partial \vec{q}$ refers to the \emph{virtual work} of potential energies. The rest of this section elaborates the kinetic and potential energy terms of Eq. (\ref{eq:govern}) for ANCF14. 

\subsection{Kinetic energy of ANCF14}
\label{subsec:kineticenergy}

Referring to Eq. (\ref{eq:position}), the velocity vector of a point P on the cross-section in the inertial (global) frame reads

\begin{equation}\label{eq:velocity}
\begin{aligned}
\frac{d \vec{r}_{\mathrm{P}}}{dt}=\dot{\vec{r}}_{\mathrm{P}} = \dot{\vec{r}}+ \dot{\mathbf{R}}\vec{h}
\end{aligned}
\end{equation}

\noindent Using this equation, the kinetic energy of an ANCF14 element  can be formulated as

\begin{equation}\label{eq:kinetic}
\begin{aligned}
T_{\text{ANCF14}} & = \frac{1}{2} \int\limits_V \rho \: \left( \dot{\vec{r}}^{\mathrm{T}}_{\mathrm{P}} \dot{\vec{r}}_{\mathrm{P}} \right) dV = \frac{1}{2} \int\limits_V \rho \left( \dot{\vec{r}}^{\mathrm{T}} \dot{\vec{r}} + 2 \dot{\vec{r}}^{\mathrm{T}} \dot{\mathbf{R}} \vec{h} + \left( \dot{\mathbf{R}} \vec{h} \right)^{\mathrm{T}} \left( \dot{\mathbf{R}} \vec{h} \right)\right) dV
\end{aligned}
\end{equation}

\noindent Knowing that $\dot{\mathbf{R}} = \begin{bmatrix} \dot{\vec{t}} \quad & \quad \dot{\vec{y}} \quad & \quad \dot{\vec{z}} \end{bmatrix}$ and $\vec{h} = \begin{bmatrix} 0 \quad & \quad \bar{y} \quad & \quad \bar{z} \end{bmatrix}^{\mathrm{T}}$, Eq. (\ref{eq:kinetic}) becomes

\begin{equation}\label{eq:kineticsimple}
\begin{aligned}
T_{\text{ANCF14}} &= \frac{1}{2} \int\limits_V \rho \left( \dot{\vec{r}}^{\mathrm{T}} \dot{\vec{r}} + 2 \dot{\vec{r}}^{\mathrm{T}} \dot{\mathbf{R}} \vec{h} +  \bar{y}^2 \dot{\vec{y}}^{\mathrm{T}} \dot{\vec{y}} + 2 \bar{y} \bar{z} \dot{\vec{y}}^{\mathrm{T}} \dot{\vec{z}} + \bar{z}^2 \dot{\vec{z}}^{\mathrm{T}} \dot{\vec{z}} \right) dV
\end{aligned}
\end{equation}

\noindent As the beam center-line passes through the cross-section's centroid, applying Eq. (\ref{eq:materialfrombishop}) to Eq. (\ref{eq:kineticsimple}) results in

\begin{equation}\label{eq:kineticfinal}
\begin{aligned}
T_{\text{ANCF14}} &= \frac{1}{2} \int\limits_V \rho \left( \dot{\vec{r}}^{\mathrm{T}} \dot{\vec{r}} + \dot{\theta}^2 \vec{h}^{\mathrm{T}} \vec{h} \right) dV \coloneqq T_{\text{translational}} + T_{\text{rotational}}
\end{aligned}
\end{equation}

\noindent Therefore, the kinetic energy consists of translational and rotational energy terms associated with the center-line motion and cross-sectional rotation, respectively. For an ANCF14 beam element with length $l$, uniform cross-sectional area $A$, uniform second polar moment of area $J$ and density $\rho$, incorporating Eq. (\ref{eq:position}) and \ref{eq:rotationinterpolation} into Eq. (\ref{eq:kineticfinal}) leads to

\begin{equation}\label{eq:finalkinetic}
\begin{aligned}
T_{\text{ANCF14}} &= \frac{\rho A}{2} \dot{\vec{q}}^\mathrm{T} \left[  \int_0^l \mathbf{S}^\mathrm{T}\mathbf{S} dx \right]\dot{\vec{q}} \\
&+ \frac{\rho J}{2} \begin{bmatrix}
\dot{\theta}_i & \dot{\theta}_j
\end{bmatrix} \left[  \int_0^l \begin{bmatrix}
(1-\frac{x}{l})^2 & \frac{x}{l}(1-\frac{x}{l}) \\
\frac{x}{l}(1-\frac{x}{l}) & \frac{x}{l}^2 
\end{bmatrix} dx \right]\begin{Bmatrix}
\dot{\theta}_i \\ \dot{\theta}_j
\end{Bmatrix}\\
& \coloneqq \frac{1}{2} \dot{\vec{q}}^\mathrm{T} \: \mathbf{M}_{\text{ANCF14}} \: \dot{\vec{q}}
\end{aligned}
\end{equation}

\noindent with $\mathbf{M}_{\text{ANCF14}}$ denoting the positive-definite \emph{mass matrix} of ANCF14 stated as

\begin{equation}\label{eq:massmatrix}
\mathbf{M}_{\text{ANCF14}} = \frac{\rho l}{420} \begin{bmatrix}
156A \mathbf{I}_{3} \\[0.8em]
22Al \mathbf{I}_{3} & 4Al^2 \mathbf{I}_{3} \\[0.8em]
\vec{0}_3^{\mathrm{T}} & \vec{0}_3^{\mathrm{T}} & 140J & & sym. \\[0.8em]
54A \mathbf{I}_{3} & 13Al \mathbf{I}_{3} & 0 & 156A \mathbf{I}_{3} \\[0.8em]
-13Al \mathbf{I}_{3} & -3Al^2 \mathbf{I}_{3} & 0 & -22Al \mathbf{I}_{3} & 4Al^2 \mathbf{I}_{3} \\[0.8em]
\vec{0}_3^{\mathrm{T}} & \vec{0}_3^{\mathrm{T}} & 70J & \vec{0}_3^{\mathrm{T}} & \vec{0}_3^{\mathrm{T}} & 140J\\
\end{bmatrix}
\end{equation}

\noindent where $\mathbf{I}_{3}$ is the $3 \times 3$ identity matrix. There is no time-dependent terms in Eq. (\ref{eq:massmatrix}) and thus the mass matrix of ANCF14 is constant and needs to be computed only once at the beginning of the simulation. For a multibody system composed of only beams discretized by ANCF14 elements and joints between them, Eq. (\ref{eq:govern}) simplifies to

\begin{equation}\label{eq:simplegovern}
\begin{aligned}
\mathbf{M}\ddot{\vec{q}} + \frac{\partial U}{\partial \vec{q}} + \frac{\partial \vec{g}}{\partial \vec{q}}^{\mathrm{T}} \vec{\lambda} &= \vec{0} \\
\vec{g}&=\vec{0}
\end{aligned}
\end{equation}

\noindent where $\mathbf{M}$ is the mass matrix of the entire multibody system.

\subsection{Potential energy of ANCF14}
\label{subsec:potentialenergy}

The potential energy of an ANCF14 element is due to gravity and its elastic deformations

\begin{equation}\label{eq:deltaU}
U_{\text{ANCF14}} \coloneqq U_{\text{gravity}} + U_{\text{elastic}}
\end{equation}

\noindent The gravity potential energy is the same as that of the standard ANCF-based beam elements

\begin{equation}\label{eq:gravbeam}
\begin{aligned}
U_{\text{gravity}} = - \int\limits_V \rho \: \vec{r}^{\mathrm{T}} \vec{\nu} \: dV = - \int\limits_V \rho \: \left( \mathbf{S} \vec{q} \right)^{\mathrm{T}} \vec{\nu} \: dV 
\end{aligned}
\end{equation}

\noindent with $\vec{\nu}$ as the gravity vector. The virtual work of gravity for an ANCF14 therefore becomes

\begin{equation}\label{eq:gengravbeam}
\begin{aligned}
\frac{\partial U_{\text{gravity}}}{\partial \vec{q}} &= - \int\limits_V \rho \mathbf{S}^{\mathrm{T}} \vec{\nu} \: dV  = - \frac{1}{12} \rho A l \begin{bmatrix}
6\mathbf{I}_{3} \quad & \quad l\mathbf{I}_{3} \quad & \quad \vec{0}_3 \quad & \quad 6 \mathbf{I}_{3} \quad & \quad -l\mathbf{I}_{3} \quad & \quad \vec{0}_3
\end{bmatrix}^{\mathrm{T}} \vec{\nu}
\end{aligned}
\end{equation}

\noindent In the next section, the derivation of elastic energy for isotropic ANCF14 beam elements considering the general case of large deformations is presented.

\subsection{Elastic energy of ANCF14}
\label{subsec:elasticenergy}

Using the \emph{second Piola-Kirchhoff stress} tensor $\mathbf{\boldsymbol\sigma}$ and the \emph{Green-Lagrange strain} tensor $\mathbf{E}$, the elastic energy of a deformable body is formulated as

\begin{equation}\label{eq:elastic}
U_{\text{elastic}} = \frac{1}{2}\int\limits_V \mathbf{\boldsymbol\sigma}:\mathbf{E} \: dV = \frac{1}{2}\int\limits_V \mathbf{C}\mathbf{E}:\mathbf{E} \: dV
\end{equation}

\noindent where $\mathbf{C}$ is the fourth-order \emph{stiffness} tensor. For ANCF14 a simple expression for $U_{\text{elastic}}$ can be obtained as follows. In terms of the deformation gradient tensor $\mathbf{F}$, the tensor $\mathbf{E}$ is given by

\begin{equation}\label{eq:green}
\mathbf{E} = \frac{1}{2} \left(\mathbf{F}^{\mathrm{T}} \mathbf{F} - \mathbf{I}\right)
\end{equation}

\noindent Noting the material frame $\mathbf{R} = \begin{bmatrix} \vec{t} \quad & \quad \vec{y} \quad & \quad \vec{z} \end{bmatrix}$ and considering Eq. (\ref{eq:position}), $\mathbf{F}$ for a point $\mathrm{P}$ on the cross-section by definition is

\begin{equation}\label{eq:defgrad}
\begin{aligned}
\mathbf{F}(x,\bar{y}, \bar{z}) & = \begin{bmatrix}
\partial \vec{r}_{\mathrm{P}} / \partial x \quad & \quad \partial \vec{r}_{\mathrm{P}} / \partial \bar{y} \quad & \quad \partial \vec{r}_{\mathrm{P}} / \partial \bar{z}
\end{bmatrix} \\
&= \begin{bmatrix}
\partial \vec{r} / \partial x + \left(\partial \mathbf{R}/ \partial x\right) \vec{h} \quad & \quad \mathbf{R} \left(\partial \vec{h}/ \partial \bar{y} \right) \quad & \quad \mathbf{R} \left( \partial \vec{h} / \partial \bar{z} \right) \end{bmatrix} \\
&= \begin{bmatrix}
\vec{r} ' + \mathbf{R}'\vec{h} \quad & \quad \vec{y} \quad & \quad \vec{z}
\end{bmatrix} \\
\end{aligned}
\end{equation}

\noindent where $x$ is the cross-section's distance along the center-line from Node $i$. Left-multiplying $\mathbf{F}$ by $\mathbf{R}^{\mathrm{T}}$ and subtracting $\mathbf{I}$ from it leads to tensor $\mathbf{D}$ providing a measure of the beam's deformation

\begin{equation}\label{eq:defmeasure}
\begin{aligned}
\mathbf{D}(x,\bar{y}, \bar{z}) &= \mathbf{R}^{\mathrm{T}}\mathbf{F}-\mathbf{I} \coloneqq \begin{bmatrix}
\vec{d}_1 \quad & \quad \vec{d}_2 \quad & \quad \vec{d}_3
\end{bmatrix} \\
&= \begin{bmatrix}
\mathbf{R}^{\mathrm{T}} \left(\vec{r} ' + \mathbf{R}'\vec{h} \right)-\vec{X} \quad & \quad \vec{0}_3 \quad & \quad \vec{0}_3
\end{bmatrix} \\
&= \begin{bmatrix}
\mathbf{R}^{\mathrm{T}} \left(\vec{r} ' - \vec{t} \: \right) + \mathbf{R}^{\mathrm{T}} \mathbf{R}'\vec{h} \quad & \quad \vec{0}_3 \quad & \quad \vec{0}_3
\end{bmatrix}
\end{aligned}
\end{equation}

\noindent in which $\vec{X} \coloneqq \begin{bmatrix} 1 \: & \: 0 \: & \: 0 \end{bmatrix}^{\mathrm{T}}$. Note that $\mathbf{D}$ is different from the material (Lagrangian) displacement gradient tensor defined as $\mathbf{F} - \mathbf{I}$. Also, $\mathbf{D}$ is not symmetric and so not the same as the Biot strain tensor as well. The second and third columns of $\mathbf{D}$ correspond to the cross-section's deformation along the $\vec{y}$ and $\vec{z}$ axes of the material frame. Having $\vec{d}_2 = \vec{d}_3 = \vec{0}_3$ implies that the cross-section remains orthogonal to the center-line and rotates rigidly about it, complying with the Euler-Bernoulli beam theory assumed initially.  Knowing $\vec{h} = \begin{bmatrix} 0 \: \: &  \bar{y} \: \: & \bar{z}\end{bmatrix}^{\mathrm{T}}$, $\vec{t} = \begin{bmatrix} t_0 \: \: & t_1 \: \: &  t_2 \end{bmatrix}^{\mathrm{T}}$,  $\vec{y} = \begin{bmatrix} y_0 \: \: &  y_1 \: \: &  y_2 \end{bmatrix}^{\mathrm{T}}$ and $\vec{z} = \begin{bmatrix} z_0 \: \: &  z_1 \: \: &  z_2 \end{bmatrix}^{\mathrm{T}}$, and based on Eq. (\ref{eq:materialdarboux}) $\gamma_1 = \vec{t} \cdot \vec{y}'$, $\gamma_2 = \vec{t} \cdot \vec{z}'$ and $\tau_{\text{m}} = \vec{y} \cdot \vec{z}'$, vector $\vec{d}_1$ can be re-formulated to

\begin{equation}\label{eq:d1}
\begin{aligned}
\vec{d}_1 = \mathbf{R}^{\mathrm{T}} \left(\vec{r} ' - \vec{t} \: \right) + \mathbf{R}^{\mathrm{T}} \mathbf{R}'\vec{h} &= \begin{Bmatrix}
\| \vec{r} ' \|-1 \\ 0 \\ 0
\end{Bmatrix} + \begin{bmatrix} t_0 \: \: & t_1 \: \: & t_2 \\[0.2em]
y_0 \: \: & y_1 \: \: & y_2 \\[0.2em]
z_0 \: \:  & z_1 \: \: & z_2
\end{bmatrix} \begin{bmatrix}
t_0' \: \: & y_0' \: \: & z_0'  \\[0.2em]
t_1' \: \: & y_1' \: \: & z_1' \\[0.2em]
t_2' \: \: & y_2' \: \: & z_2'
\end{bmatrix} \begin{Bmatrix} 0 \\ \bar{y} \\ \bar{z}  \end{Bmatrix} \\
&= \begin{Bmatrix}
\| \vec{r} ' \|-1 \\ 0 \\ 0
\end{Bmatrix} + \begin{bmatrix}
0 \quad \: & \vec{t} \cdot \vec{y}' \quad \: & \vec{t} \cdot \vec{z}' \\[0.2em]
\vec{y} \cdot \vec{t}' \quad \: & 0 \quad \: & \vec{y} \cdot \vec{z}' \\[0.2em]
\vec{z} \cdot \vec{t}' \quad \: & \vec{z} \cdot \vec{y}' \quad \: & 0
\end{bmatrix} \begin{Bmatrix} 0 \\ \bar{y} \\ \bar{z}  \end{Bmatrix} \\
&= \begin{Bmatrix}
\| \vec{r} ' \|-1 \\ 0 \\ 0
\end{Bmatrix}  + \begin{Bmatrix}
\gamma_1\bar{y}+\gamma_2\bar{z} \\ \tau_m \bar{z} \\ -\tau_m \bar{y}
\end{Bmatrix}
\end{aligned}
\end{equation}

\noindent The first part of $\vec{d}_1$ describes the longitudinal deformations and the second part characterizes bending and torsional deformations. Utilizing the derived tensor $\mathbf{D}$, the Green-Lagrange strain becomes

\begin{equation}\label{eq:finalgreen}
\mathbf{E} = \frac{1}{2} \left(\mathbf{F}^{\mathrm{T}} \mathbf{F} - \mathbf{I}\right) = \frac{1}{2} \left(\mathbf{D} + \mathbf{D}^{\mathrm{T}} + \mathbf{D}^{\mathrm{T}} \mathbf{D} \right)
\end{equation}

\noindent Upon putting Eq. (\ref{eq:defmeasure}) and Eq. (\ref{eq:d1}) into Eq. (\ref{eq:finalgreen}), one finds

\begin{equation}\label{eq:finalgreenelems}
\begin{gathered}
\mathbf{E} \coloneqq \begin{bmatrix}
\varepsilon_{11} & \varepsilon_{12} & \varepsilon_{13} \\
\varepsilon_{12} & 0 & 0 \\
\varepsilon_{13} & 0 & 0
\end{bmatrix} \\
\varepsilon_{11}=\frac{1}{2} \left( \left(\|\vec{r}'\|-1+ \gamma_1\bar{y}+\gamma_2\bar{z} \right)^2 + \tau_m^2\bar{y}^2+\tau_m^2\bar{z}^2 +2\left(\|\vec{r}' \| -1 + \gamma_1\bar{y}+\gamma_2\bar{z} \right) \right) \\
\varepsilon_{12} = \frac{1}{2}\tau_m \bar{z} \quad , \quad \varepsilon_{13} = -\frac{1}{2} \tau_m\bar{y} 
\end{gathered}
\end{equation}

The expression for $\varepsilon_{11}$ suggests a coupling between the longitudinal, transverse bending and torsional deformations. This dependence arises from the quadratic term $ \mathbf{D}^{\mathrm{T}} \mathbf{D}$ in Eq. (\ref{eq:finalgreen}). It is well-known that if the deformations are small and within the linear elastic regime, this term can be neglected \cite{shabana2020dynamics}, and $\varepsilon_{11}$ simplifies to $\varepsilon_{11}^{\mathrm{small}}$ as

\begin{equation}\label{eq:simplee11}
\varepsilon_{11}^{\mathrm{small}} = \| \vec{r} ' \| -1 + \gamma_1 \bar{y} + \gamma_2 \bar{z} 
\end{equation}

\noindent which results in

\begin{equation}\label{eq:e11basedonsmall}
\varepsilon_{11}=\varepsilon_{11}^{\mathrm{small}}+\frac{1}{2} \left(\left( \varepsilon_{11}^{\mathrm{small}}\right)^2+\tau_m^2 \left(\bar{y}^2+\bar{z}^2 \right)\right)
\end{equation}

\noindent Putting Eq. (\ref{eq:finalgreenelems}) into Eq. (\ref{eq:elastic}), the total elastic energy of an ANCF14 element reads

\begin{equation}\label{eq:totalelastic}
U_{\text{elastic}} = \frac{1}{2}\int\limits_V E\varepsilon_{11}^2+4 G\varepsilon_{12}^2+4 G\varepsilon_{13}^2 \: dV
\end{equation}

\noindent where $E$ is the Young's modulus and $G$ denotes the shear modulus. Provided these conditions for the cross-section: i) it is uniform along the element ii) the origin of the material frame coincides with its centroid, and iii) $\vec{y}$ and $\vec{z}$ are symmetry axes of it, the elastic energy for small strains reduces to

\begin{equation}\label{eq:totalelasticsmalldef}
\begin{aligned}
U_{\text{elastic}}\Big|_\mathrm{small} &= \frac{EA}{2} \int_0^l \left(\| \vec{r} ' \| -1\right)^2  \: dx \\
&+ \frac{EI_z}{2} \int_0^l \gamma_1^2  \: dx + \frac{EI_y}{2} \int_0^l \gamma_2^2  \: dx + \frac{GJ_t}{2} \int_0^l \tau_m^2  \: dx
\end{aligned}
\end{equation}

\noindent in which $I_y$ and $I_z$ are the cross-section's second moments of area about $\vec{y}$ and $\vec{z}$; and $J_t$ is the torsional constant of the beam's cross-section. Under the same conditions set for deriving Eq. (\ref{eq:totalelasticsmalldef}) and assuming the fourth moments of area

\begin{equation}
\int\limits_A \bar{y}^4 \: dA \approx 0, \: \int\limits_A \bar{z}^4 \: dA \approx 0, \int\limits_A \bar{y}^2 \bar{z}^2 \: dA \approx 0
\end{equation}

\noindent the elastic  potential energy for the general case of large deformations yields 

\begin{equation}\label{eq:totalelasticdef}
\begin{aligned}
U_{\text{elastic}} &= \frac{EA}{8} \int_0^l \left(\| \vec{r} ' \|^2 -1\right)^2  \: dx \\
&+ \frac{EI_z}{4} \int_0^l \gamma_1^2 \left( 3 \| \vec{r}' \|^2 - 1 \right) \: dx \\
&+ \frac{EI_y}{4} \int_0^l \gamma_2^2 \left( 3 \| \vec{r}' \|^2 - 1 \right) \: dx \\
&+ \frac{GJ_t}{2} \int_0^l \tau_m^2  \: dx + \frac{EJ_t}{4} \int_0^l \tau_m^2 \left( \| \vec{r}' \|^2- 1 \right) \: dx \\
\end{aligned}
\end{equation}

\noindent This equation shows that in large deformation cases, the longitudinal, bending and torsional deformations are strongly coupled. The derivation of Eq. (\ref{eq:totalelasticdef}) is detailed in Appendix \ref{appen:ularge}. Both Eq. (\ref{eq:totalelasticsmalldef}) and Eq. (\ref{eq:totalelasticdef}) are in the so-called \emph{structural mechanics-based formulation} \cite{nachbagauer2013structural}, \cite{simo1985finite}.

\subsection{Virtual work of elastic energy}
\label{subsec:vwelastic}

The virtual work of elastic energies (i.e. the internal elastic forces) is the derivative of the elastic energy with respect to the element's state variables. Thus, employing Eq. (\ref{eq:e11basedonsmall}) and Eq. (\ref{eq:totalelastic}) yields

\begin{equation}\label{eq:vwder}
\begin{aligned}
\frac{\partial U_{\text{elastic}}}{\partial \vec{q}} &= \int\limits_V \left( E\varepsilon_{11}\frac{\partial \varepsilon_{11}}{\partial \vec{q}}+4 G\varepsilon_{12}\frac{\partial \varepsilon_{12}}{\partial \vec{q}}+4G\varepsilon_{13}\frac{\partial \varepsilon_{13}}{\partial \vec{q}} \right) \: dV \\
&= E\int\limits_V \varepsilon_{11} \left( (1+\varepsilon_{11}^{\mathrm{small}})\frac{\partial \varepsilon_{11}^{\mathrm{small}}}{\partial \vec{q}} + \tau_m \left(\bar{y}^2+\bar{z}^2 \right) \frac{\partial \tau_m}{\partial \vec{q}} \right) \: dV \\
&+G\int\limits_V \left( \tau_m \left(\bar{y}^2+\bar{z}^2 \right) \frac{\partial \tau_m}{\partial \vec{q}}\right) \: dV
\end{aligned}
\end{equation}

\noindent Based on Eq. (\ref{eq:simplee11})

\begin{equation}\label{eq:dsmalldq}
\begin{aligned}
\frac{\partial \varepsilon_{11}^{\mathrm{small}}}{\partial \vec{q}} &= \frac{1}{\| \vec{r}' \|} \vec{r}'^{\mathrm{T}} \frac{\partial \vec{r}'}{\partial \vec{q}} + \bar{y} \frac{\partial \gamma_1}{\partial \vec{q}} + \bar{z} \frac{\partial \gamma_2}{\partial \vec{q}}
\end{aligned}
\end{equation}

\noindent and using the assumptions made for Eq. (\ref{eq:totalelasticsmalldef}), Eq. (\ref{eq:vwder}) for small strains reduces to

\begin{equation}\label{eq:vwdersmall}
\begin{aligned}
\frac{\partial U_{\text{elastic}}}{\partial \vec{q}}\Big|_\mathrm{small} &= E\int\limits_V \left( \varepsilon_{11}^{\mathrm{small}}\frac{\partial \varepsilon_{11}^{\mathrm{small}}}{\partial \vec{q}} \right) \: dV +G\int\limits_V \left( \tau_m \left(\bar{y}^2+\bar{z}^2 \right) \frac{\partial \tau_m}{\partial \vec{q}}\right) \: dV \\
&= EA \int_0^l \left[ \left( 1 - \frac{1}{\| \vec{r} ' \|} \right) \vec{r}'^{\mathrm{T}} \frac{\partial \vec{r}'}{\partial \vec{q}} \right] \: dx \\
&+E I_z \int_0^l \gamma_1  \frac{\partial \gamma_1}{\partial \vec{q}} \: dx + E I_y \int_0^l \gamma_2  \frac{\partial \gamma_2}{\partial \vec{q}} \: dx + GJ_t \int_0^l \tau_m  \frac{\partial \tau_m}{\partial \vec{q}} \: dx 
\end{aligned}
\end{equation}

\noindent in which the terms involving $EA$, $E I_z$, $E I_y$ and $G J_t$ account for the virtual works of elastic energies (internal forces) respectively associated with axial, bending about local $\vec{y}$ axis, bending about local $\vec{z}$ axis and torsional deformations. For large deformations Eq. (\ref{eq:totalelasticdef}), $\partial U_{\text{elastic}} / \partial \vec{q}$ can be calculate in a similar manner as

\begin{equation}\label{eq:vwderlarge}
\begin{aligned}
\frac{\partial U_{\text{elastic}}}{\partial \vec{q}} &= \frac{EA}{2} \int_0^l \left[ \left( \| \vec{r} ' \|^2 -1 \right) \vec{r}'^{\mathrm{T}} \frac{\partial \vec{r}'}{\partial \vec{q}} \right] \: dx \\
&+ \frac{E I_z}{2} \int_0^l \left[ \left( 3\| \vec{r} ' \|^2 -1  \right) \gamma_1 \frac{\partial \gamma_1}{\partial \vec{q}} + 3\gamma_1^2 \vec{r}'^{\mathrm{T}} \frac{\partial \vec{r}'}{\partial \vec{q}} \right] \: dx \\
&+ \frac{E I_y}{2} \int_0^l \left[ \left( 3\| \vec{r} ' \|^2 -1  \right) \gamma_2 \frac{\partial \gamma_2}{\partial \vec{q}} + 3\gamma_2^2 \vec{r}'^{\mathrm{T}} \frac{\partial \vec{r}'}{\partial \vec{q}} \right] \: dx \\
&+ \frac{E J_t}{2} \int_0^l \left[ \left( \| \vec{r} ' \|^2 -1  \right) \tau_{\text{m}} \frac{\partial \tau_{\text{m}}}{\partial \vec{q}} + \tau_{\text{m}}^2 \vec{r}'^{\mathrm{T}} \frac{\partial \vec{r}'}{\partial \vec{q}} \right] \: dx \\
&+ G J_t \int_0^l \tau_{\text{m}} \frac{\partial \tau_{\text{m}}}{\partial \vec{q}} \: dx
\end{aligned}
\end{equation}


\noindent which again demonstrates the strong coupling between the longitudinal, bending and torsional deformations, as well as their associated internal forces, in large deformations. 

Both of Eq. (\ref{eq:vwdersmall}) and Eq. (\ref{eq:vwderlarge}) are nonlinear in terms of $\vec{q}$, as is the case for $\partial U_{\text{elastic}} / \partial \vec{q}$ of other ANCF-based beam elements, and numerical integration methods (e.g. Gaussian quadrature) can be employed to compute them. To do so, however, $\partial \vec{r}' /\partial \vec{q}$, $\partial \gamma_1/\partial \vec{q}$, $\partial \gamma_2/\partial \vec{q}$ and $\partial \tau_m/\partial \vec{q}$ at the integration points are required. This can be accomplished utilizing the definition of the material frame provided in Section \ref{sec:kinematics}. For the sake of brevity, the details are laid out in Appendix \ref{appen:derivative}. Once the virtual work of elastic energies is known, adopting an appropriate time-stepping technique (e.g. \emph{Runge-Kutta}, \emph{geometric variational integrators} \cite{leyendecker2008variational}, \emph{Newmark}) solves the equations of motion in Eq. (\ref{eq:simplegovern}).

\section{Numerical examples}
\label{sec:examples}

In order to validate the proposed beam element four numerical examples are provided. As capturing the torsional deformations is one the main promises of this proposal, examples in which torsion plays a major role are selected. An Autodesk's proprietery multibody static/dynamic library called Momentum has been used to run the numerical experiments. To solve the equations of motion, a geometric variational integrator introduced by Leyendecker et al. \cite{leyendecker2008variational}, also detailed in \cite{ebrahimi201982}, is employed. A Newton-Raphson scheme is used to solve the nonlinear equations that arise during the solution process.

To show the accuracy of ANCF14 results, they are compared against the theoretical result in the first example and against the numerical results of different beam formulations provided in \cite{bauchau2016validation} and \cite{yakoub2001three} for the other three benchmarks. In \cite{bauchau2016validation} a set of standard numerical tests for validating beam finite elements in multibody dynamic problems are provided. In that study, the performance of different beam models \cite{cardona1988beam, ghiringhelli2000multibody, bauchau2001modeling, betsch2002frame, lens2008nonlinear, jonker2013geometrically, nachbagauer2013structural, sonneville2014geometrically} are reported in the presented tests which serves as a proven reference to determine the accuracy of the results produced by ANCF14.

To extract the data from the plots presented in \cite{bauchau2016validation}, an open-source tool called WebPlotDigitizer \cite{rohatgi2011webplotdigitizer} is used. As analyzing the computational efficiency of different beam elements strongly depends on their implementation from the programming perspective and the machine used to run the experiments, it does not seem plausible to compare their computational efficiency utilizing their simulation time reported in other studies. Nevertheless, to demonstrate ANCF14's efficiency, we have implemented the two-noded version of the 3D ANCF element with 24 DOF introduced in \cite{yakoub2001three}, referenced hereafter as ANCF24, and their simulation times are compared against each other. Note that ANCF24 can also account for the cross-sectional shear deformations and the comparison of its computational efficiency with that of ANCF14 in this study is solely to reflect the effect of using extra degrees of freedom in the simulation time.  The computer used to execute the numerical tests is a Windows machine with an Intel Xeon E5 CPU and 32 GB of RAM.

\subsection{Helical spring}
\label{subsec:helix}

\begin{figure*}
\centering
\includegraphics[width=0.85\textwidth]{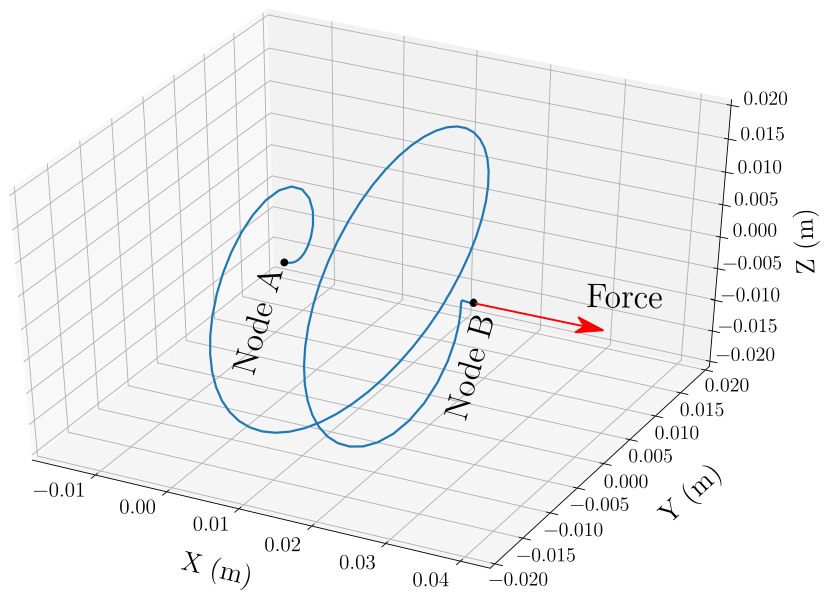}
\caption{The undeformed geometry of the spring}
\label{fig:undeformedgeom} 
\end{figure*}

The goal of the first test, inspired by \cite{fan2016accurate}, is to compare the linear stiffness $k$ of a helical spring with a circular cross-section modeled by ANCF14 against its theoretical value \cite{shigley2011shigley} approximated through 

\begin{equation}
 k \approx \frac{G \: d^4}{ 8 N_a D^3}
\end{equation}

\noindent where $G$, $d$, $D$ and $N_a$ are the shear modulus, wire diameter, mean spring diameter and number of active coils in the spring, respectively. In this example, $G=80$ GPa, $d=2$ mm, $D=40$ mm and $N_a = 1.5$, thus leading to $k \approx 1.667$ N/mm. The undeformed geometry of this spring, illustrated in Figure \ref{fig:undeformedgeom}, is defined as

\begin{equation}
\vec{r}(x) = \begin{Bmatrix}
0.05x \\
0.02 \: a(x) \cos(8 \pi x) \\
0.02 \: a(x) \sin(8 \pi x)
\end{Bmatrix}
\end{equation}

\noindent with 

\begin{equation}
a(x) = \begin{cases}
0.5 \left(1 + \tanh \left(50x-3 \right) \right)       & \quad  0 \le x < 0.15 \\
1  & \quad 0.15 \le x \le 0.35 \\
0.5 \left(1 + \tanh \left(22-50x \right) \right) & \quad 0.35 < x \le 0.5 \\
\end{cases}
\end{equation}

\begin{figure*}
\centering
\includegraphics[width=0.9\textwidth]{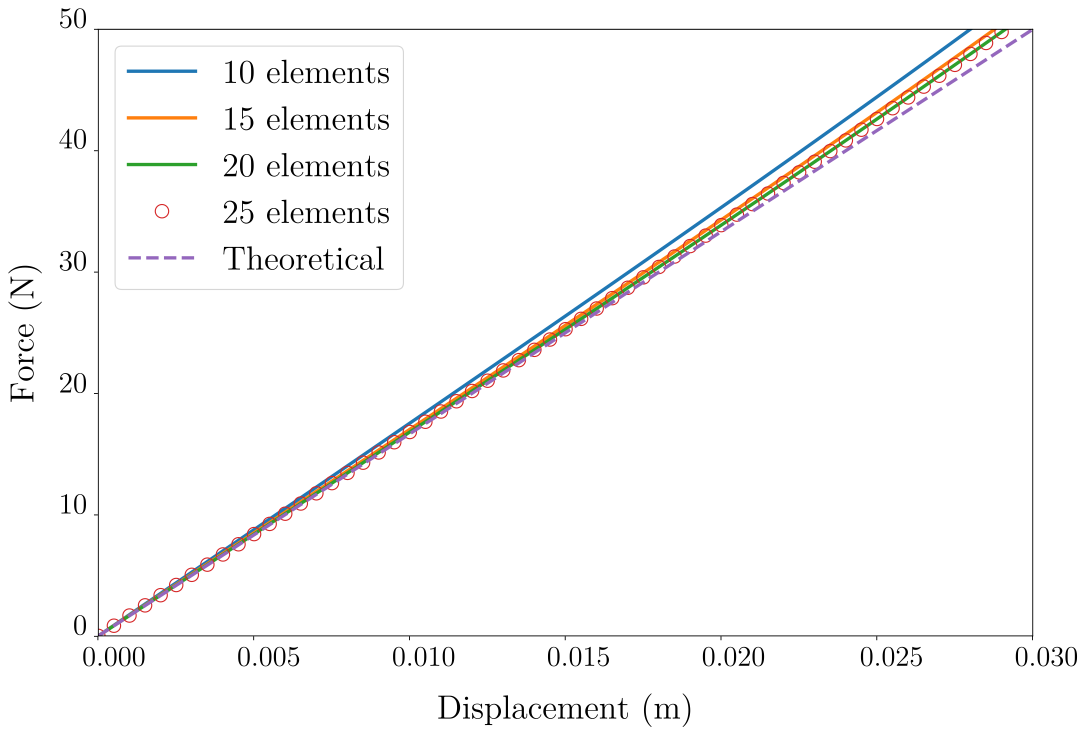}
\caption{The force-displacement diagram of Node B}
\label{fig:forcedisp} 
\end{figure*}

The spring is connected to the ground by a spherical joint at Node A and is pulled along the positive $\textrm{X}$ axis from Node B, as depicted in Figure \ref{fig:undeformedgeom}. To ensure the discretization (mesh) independency, the test is run using 10, 15, 20 and 25 straight ANCF14 elements assuming large deformations. The force-displacement diagram for Node B is presented in Figure \ref{fig:forcedisp}. Accordingly, by increasing the number of elements, the linear spring behavior converges toward the theoretical prediction. The curves for 20 and 25 elements coincide, and so the simulated behavior is shown to be mesh-independent for 20 elements onward. Running the test with a significantly more number of elements (e.g. 100) ensured no shear locking issue, as expected, since the cross-sectional shear deformations are not taken into account in ANCF14. Note that the spring behaves linearly at first and as Node B is further pulled, its force-displacement relationship becomes nonlinear. The computed linear spring stiffness using 20 elements is $1.674 \: \textrm{N/mm}$, which coincides almost perfectly with its theoretical value.

The same results are achieved using 20 ANCF24 elements, however, with 105 degrees of freedom more than 20 ANCF14 elements. This difference in the number of degrees of freedom leads to about 55 percent increase in the simulation time for ANCF24.

\begin{figure*}
\centering
\includegraphics[width=0.9\textwidth]{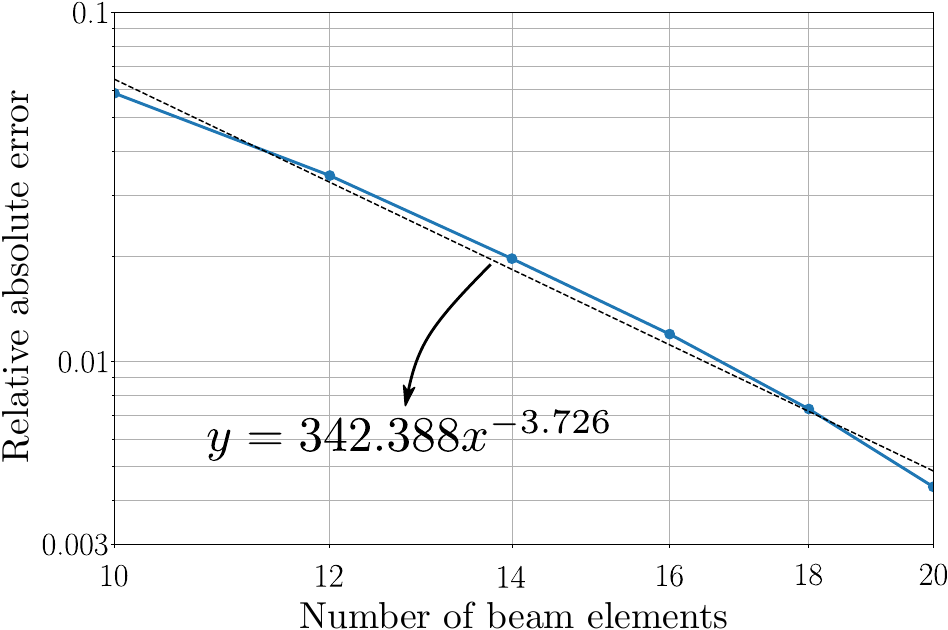}
\caption{The evolution of absolute relative error in X-displacement of Node B}
\label{fig:errorev} 
\end{figure*}

Figure \ref{fig:errorev} shows the evolution of absolute relative error in X-displacement of Node B versus the number of elements used to discretize the spring. The ground truth displacement is computed using the theoretical value of linear spring constant. As illustrated, the error reduces almost quartically ($\approx 3.7$) by increasing the number of elements.

\FloatBarrier
\subsection{Princeton beam experiment}
\label{subsec:princeton}

The Princeton beam experiment is the \emph{quasi}-static analysis of a cantilever beam with a rectangular cross-section undergoing geometric nonlinearities (large deformations and rotations) subject to a point load applied to its tip \cite{dowell1975experimental, dowell1257experimental}. Due to the complex behavior of the beam, this problem has become a standard benchmark to study the accuracy of the newly developed beam models for multibody dynamic applications. 

\begin{figure*}
\centering
\includegraphics[width=0.95\textwidth]{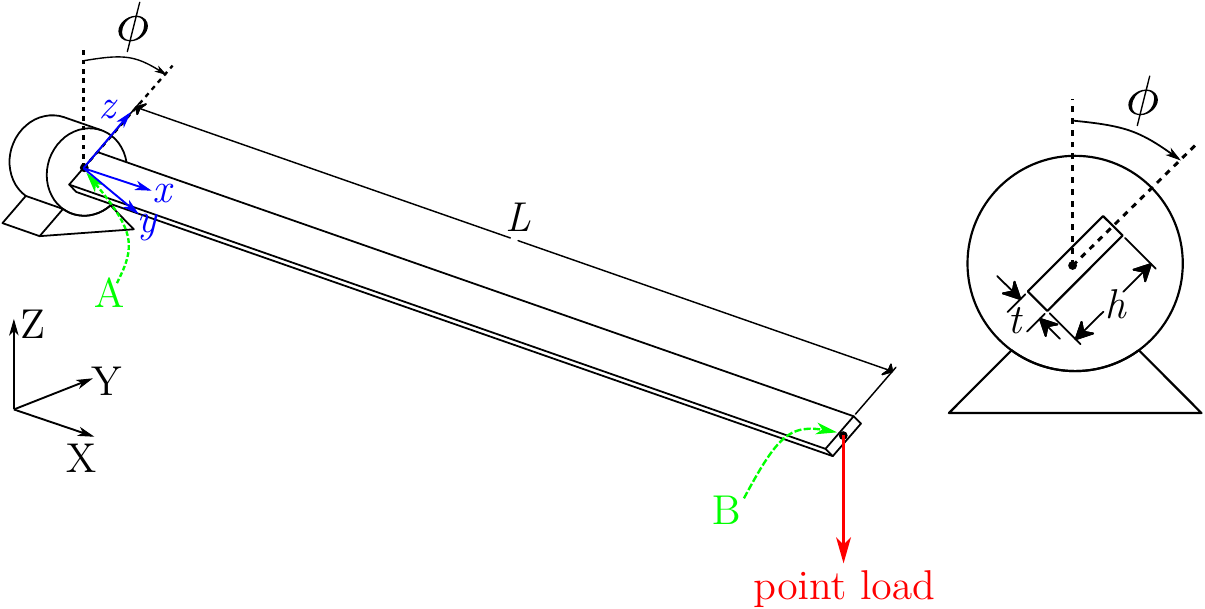}
\caption{\label{fig:setupprinceton} The set-up of the Princeton beam experiment}
\end{figure*}

The experiment set-up is illustrated in Figure \ref{fig:setupprinceton}. The beam is made of T7075 aluminum with Young's modulus $E = 71.7 \: \textrm{GPa}$ and Poisson's ratio $\nu = 0.31$. It is clamped to a rotor at Point A and a downward force is applied at Point B. The experiment studies the tip displacements along the local $y$ and $z$ axes, as well as its cross-section's twist for different angles $\phi$ measured with respect to the global Z axis. The test is run for $\phi \in [0^{\circ}, 90^{\circ}]$ with $15^{\circ}$ increments subject to three different load magnitudes $P_1 = 4.448 \:$N, $P_2 = 8.896 \:$N and $P_3 = 13.345 \:$N. The beam has a length of $L=0.5080$ m, and a rectangular cross-section with thickness $t=3.2024$ mm, height of $h=12.3770$ mm and torsion constant $J_t=113.3872 \: \mathrm{mm}^4$.

\begin{figure*}[ht]
\centering
\includegraphics[width=0.95\textwidth]{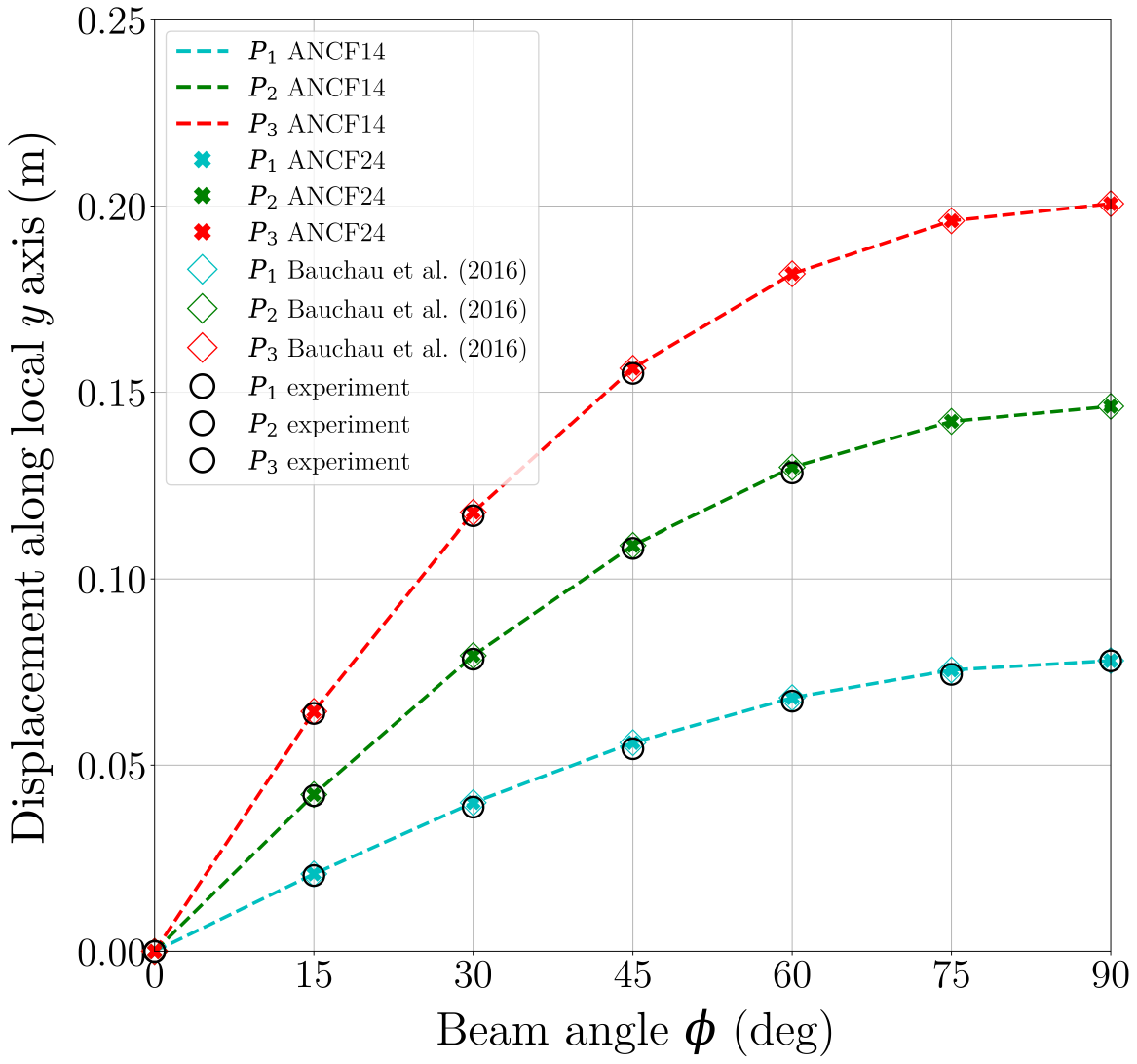}
\caption{The tip displacement along the local y axis of the Princeton experiment for different beam angles}
\label{fig:princetonu2disp} 
\end{figure*}

\begin{figure*}[ht]
\centering
\includegraphics[width=0.95\textwidth]{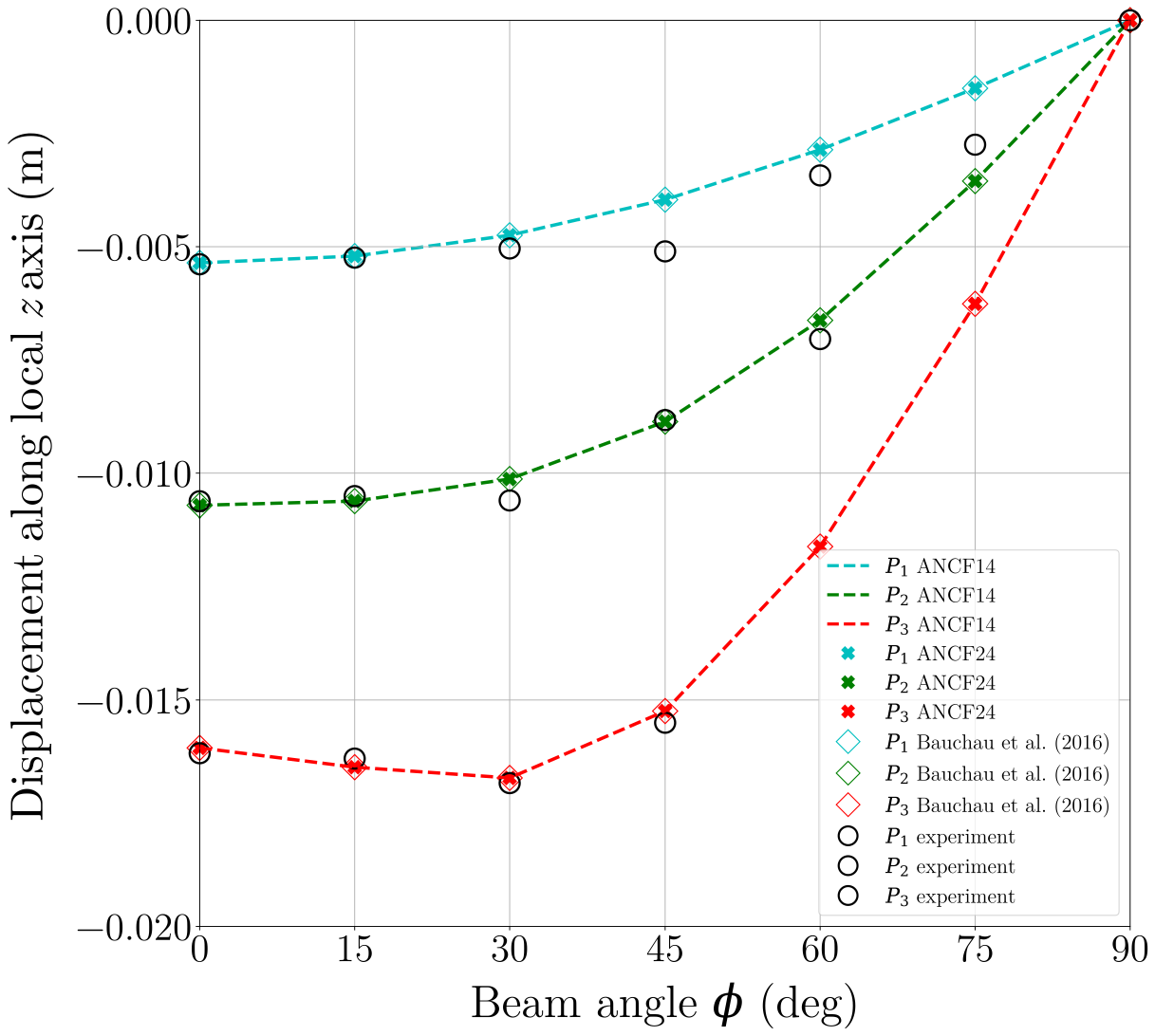}
\caption{The tip displacement along the z axis of the Princeton experiment for different beam angles}
\label{fig:princetonu3disp} 
\end{figure*}

\begin{figure*}[ht]
\centering
\includegraphics[width=0.95\textwidth]{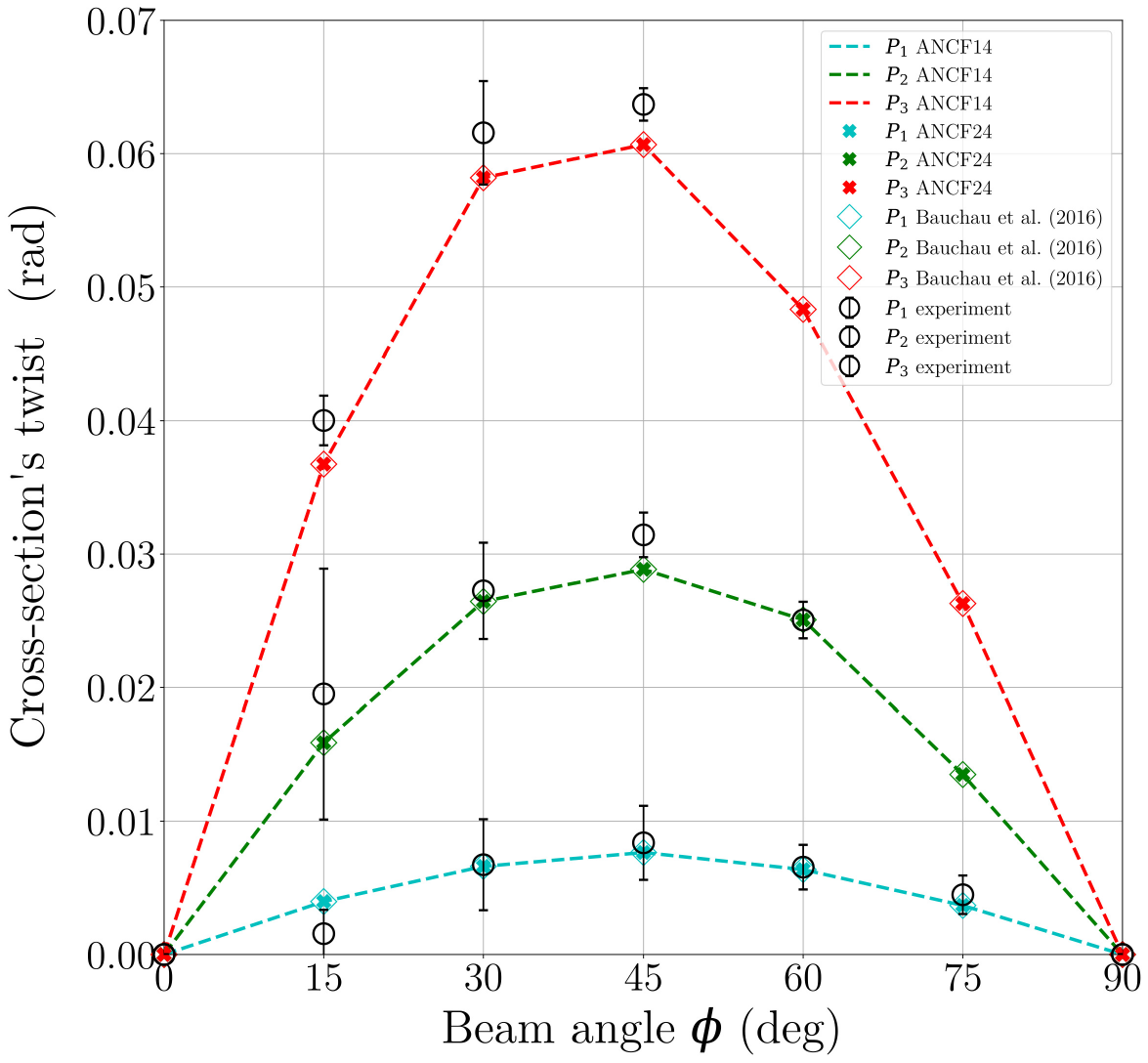}
\caption{The tip's cross-sectional twist of the Princeton experiment for different beam angles}
\label{fig:princetontwist} 
\end{figure*}

The beam is discretized using 8 ANCF14 elements with the large deformation assumption. Figures \ref{fig:princetonu2disp}-\ref{fig:princetontwist} compare our simulation results with those of several different multibody solvers (beam models) reported in \cite{bauchau2016validation} and 8 ANCF24 elements. As can been seen, the results are in excellent agreement. For this test, the simulation time for ANCF14 and ANCF24 elements are approximately 1.21 sec and 1.92 sec, respectively.

In Figures \ref{fig:princetonu2disp}-\ref{fig:princetontwist}, the experimental results are also presented. Accordingly, there is a small discrepancy between them and those of ANCF14, ANCF24 and those reported in Bauchau et al. \cite{bauchau2016validation}, particularly for the cross-section's twist. This may be attributed to one or a combination of these three reasons: 1) the linear interpolation of the twist angle in ANCF14, 2) the Euler-Bernoulli beam assumption and ignoring the cross-section's shear, especially in large deformations and 3) the difference between the properties used for the numerical analysis and those to conduct the physical testing. Considering the insignificant differences and the agreement between the outputs of ANCF14 and all the other solvers (formulations), the latter can be a plausible possibility.

\FloatBarrier
\subsection{Unbalanced rotating shaft}
\label{subsec:shaft}

\begin{figure*}[ht]
\centering
\includegraphics[width=0.90\textwidth]{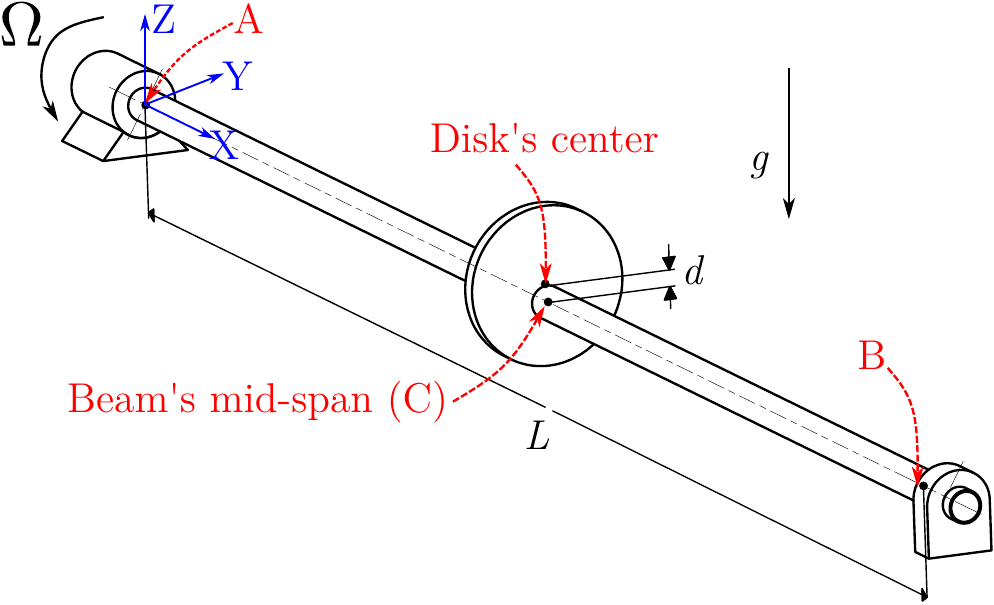}
\caption{\label{fig:setupshaft} The set-up of the unbalanced rotating shaft benchmark}
\end{figure*}

Introduced in \cite{bauchau2016validation}, this benchmark seeks to study the nonlinear dynamic behavior of an unbalanced rotating shaft subject to a downward gravity $g=9.81 \: \mathrm{m/s^2}$ and a prescribed angular velocity $\Omega(t)$ as

\begin{equation}\label{eq:angularvel}
\Omega(t) = 
\begin{cases}
0.4 \: \omega \left(1-\cos \left( 2\pi t \right) \right) & \quad  0 \le t < 0.5 \\
0.8 \: \omega & \quad 0.5 \le t < 1 \\
0.2 \: \omega \left(5-\cos \left( 4\pi (t-1) \right) \right)&  \quad  1 \le t < 1.25 \\
1.2 \: \omega & \quad 1.25 \le t \le 2.5 
\end{cases}
\end{equation}

\noindent where $\omega = 60 \:$ rad/s, close to the first bending natural frequency of the shaft (about 56.7 rad/s). The shaft is deformed initially at time $t=0$ under the effect of gravity. It is made of steel with Young's modulus $E = 210 \:$GPa, Poisson's ratio $\nu = 0.3$ and density $\rho = 7800 \: \mathrm{kg/m^3}$. The shaft is $L=6\:$m long and has a hollow circular cross-section with inner radius $r_i = 0.045\:$m and outer radius $r_o = 0.05\:$m. It is connected to the ground through a revolute joint at End A and a cylindrical joint at End B. At the shaft's mid-span, Point C, a rigid disk is welded whose center is $d = 0.05\:$m above the shaft's center-line. The disk's mass $m$ and its diagonal inertial tensor $I_{\mathrm{disk}}$ (in the inertial frame) are 70.573 kg and diag(2.0325, 1.0163, 1.0163) $\mathrm{g\:m^2}$, respectively. The problem set-up is demonstrated in Figure \ref{fig:setupshaft}.

\begin{figure*}
\centering
	\begin{subfigure}{1.0\textwidth}
        \includegraphics[width=\textwidth]{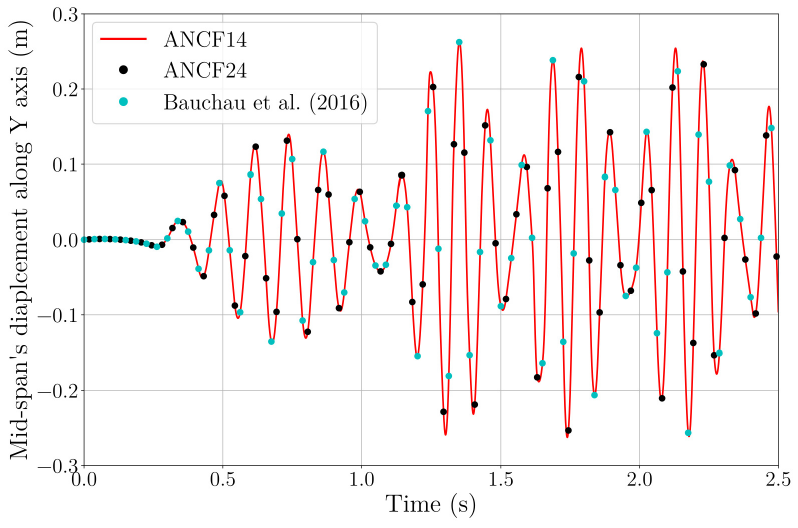}
        \caption{}
        \label{subfig:u2dispshaft}
    \end{subfigure}
    \begin{subfigure}{1.0\textwidth}
        \includegraphics[width=\textwidth]{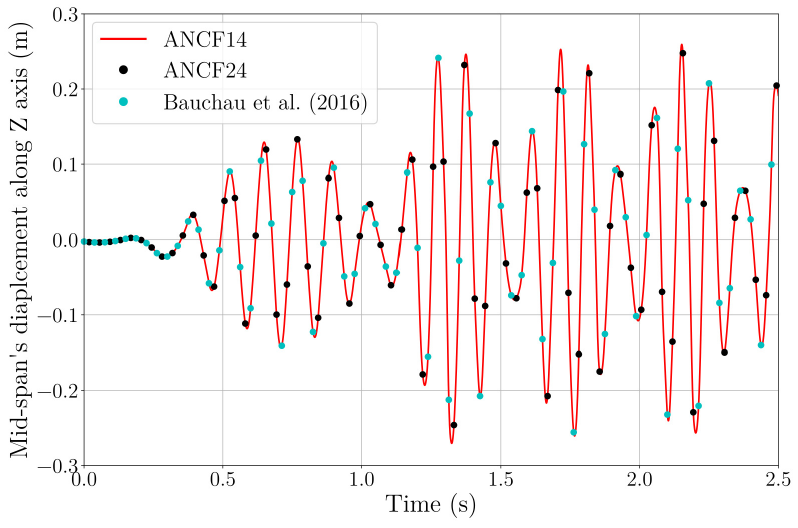}
        \caption{}
        \label{subfig:u3dispshaft}
    \end{subfigure}
    \caption{The displacement of Point C in time (a) along the Y axis (b) along the Z axis}
    \label{fig:displacements}
\end{figure*}

\begin{figure*}
\centering
\includegraphics[width=1.0\textwidth]{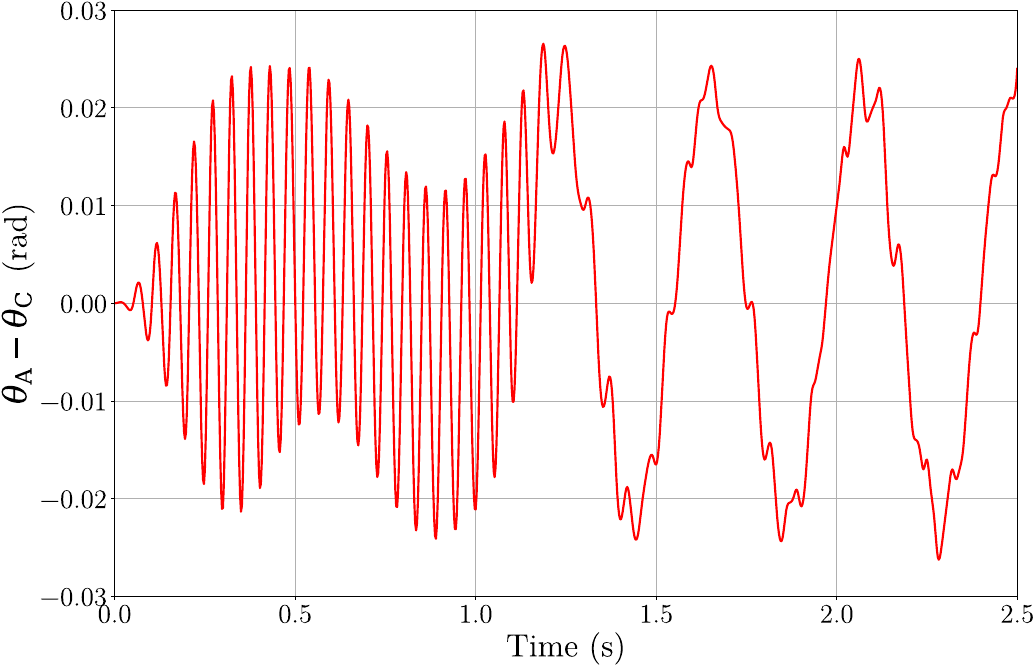}
\caption{\label{fig:twistdiff} The evolution of $\theta_{\mathrm{A}}-\theta_{\mathrm{C}}$ with time}
\end{figure*}

Figure \ref{fig:displacements} shows the displacement of Point C along the global Y and Z axes in time. In both plots, the displacement amplitudes start amplifying at around time $t \approx 1.1\:$s, when the angular velocity passes the first bending natural frequency (Eq. (\ref{eq:angularvel}) ). At this time instance, the shaft's behavior transitions from \emph{sub-critical} to \emph{super-critical}. Figure \ref{fig:twistdiff} pictures the evolution of $\theta_{\mathrm{A}}-\theta_{\mathrm{C}}$ with time, where $\theta_{\mathrm{A}}$ and $\theta_{\mathrm{C}}$ are the cross-section's rotation at End A and Point C (Eq. (\ref{eq:rotationinterpolation}) ). Clearly, its behavior changes distinctly when passing the critical $\Omega$.

In Figure \ref{fig:displacements}, the numerical results of using ANCF24 and those reported in Bauchau et al. \cite{bauchau2016validation} are presented as well. Accordingly, ANCF14 results are in great agreement with those of other formulations. For this benchmark, 6 ANCF14 elements (thus 7 nodes) with the large deformation assumption suffice to guarantee mesh independent results. As each node of ANCF14 has 7 DOF, the total number of DOF used is 49. On the other hand, using 6 ANCF24 elements leads to the total 168 DOF causing about 63 percent longer simulation time. For more complex problems, this difference could be even more significant, thus making ANCF14 more computationally advantageous.

\begin{figure*}
\centering
	\begin{subfigure}{1.0\textwidth}
        \includegraphics[width=\textwidth]{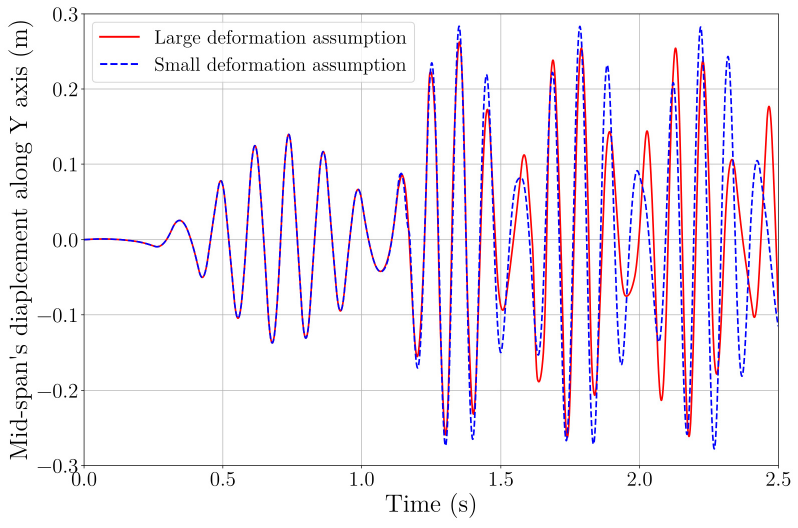}
        \caption{}
        \label{subfig:u2dispSMshaft}
    \end{subfigure}
    \begin{subfigure}{1.0\textwidth}
        \includegraphics[width=\textwidth]{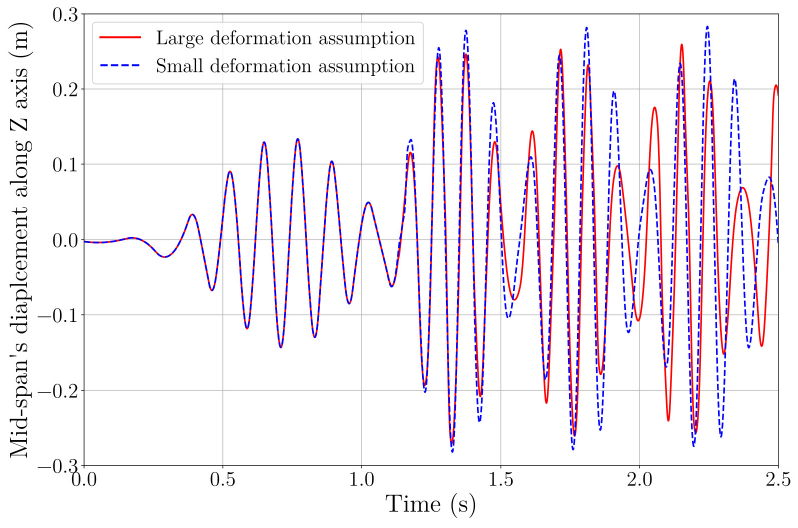}
        \caption{}
        \label{subfig:u3dispSMshaft}
    \end{subfigure}
    \caption{The displacement of Point C in time for large and small deformation assumptions (a) along the Y axis (b) along the Z}
    \label{fig:smalllarge}
\end{figure*}

In order to investigate the importance of assuming large deformations for this example, it is run using 6 ANCF14 elements with the small deformation assumption as well. Figure \ref{fig:smalllarge} compares the simulation results for the two cases. As can be seen, at the beginning where the shaft's angular velocity is below its critical value and the displacements are small, the results coincide. However, as the shaft accelerates and the displacements grow, the results start to deviate.

\FloatBarrier
\subsection{Lateral torsional buckling of a slender beam}
\label{subsec:buckling}

\begin{table}
\caption{\label{table:tablebuckling}The values of parameters in the lateral torsional buckling test }
\begin{center}
\begin{tabular}{lll}
\toprule
\textbf{Parameter} &\hspace{1em} \textbf{Value} & \textbf{Unit} \\[0.2em]
\hline
Young's Modulus $E$ &\hspace{1em} $73$ & GPa \\[0.2em]
Poisson's ration $\nu$ &\hspace{1em} 0.3 & \\[0.2em]
Density $\rho$ &\hspace{1em} $2680$ & $\mathrm{kg/m^3}$ \\[0.2em]
Beam's length $L$  &\hspace{1em} $1.0$ & m  \\[0.2em]
Height of beam's cross-section $h$ &\hspace{1em} 0.01 & m  \\[0.2em]
Width of beam's cross-section $w$ &\hspace{1em} 0.001 & m  \\[0.2em]
Beam's torsion constant $J_{t_{\mathrm{b}}}$ &\hspace{1em} $3.12 \times 10^{-8}$ &  $\textrm{m}^4$  \\[0.2em]
Link's length $L_{\mathrm{l}}$  &\hspace{1em} $0.25 \:$ & m  \\[0.2em]
Radius of link's cross-section $r_{\mathrm{l}}$ &\hspace{1em} 0.012 & m  \\[0.2em]
Link's torsion constant $J_{t_{\mathrm{l}}}$ &\hspace{1em} $3.26 \times 10^{-8}$ &  $\textrm{m}^4$  \\[0.2em]
Crank's length $L_{\mathrm{c}}$  &\hspace{1em} $0.05 \:$ & m  \\[0.2em]
Radius of crank's cross-section $r_{\mathrm{c}}$ &\hspace{1em} 0.024 & m  \\[0.2em]
Crank's torsion constant $J_{t_{\mathrm{c}}}$ &\hspace{1em} $5.21 \times 10^{-8}$ &  $\textrm{m}^4$  \\[0.2em]
Offset $d$ &\hspace{1em} $10^{-4}$ & m  \\[0.2em]
\bottomrule
\end{tabular}
\end{center}
\end{table}

\begin{figure*}
\centering
\includegraphics[width=1.0\textwidth]{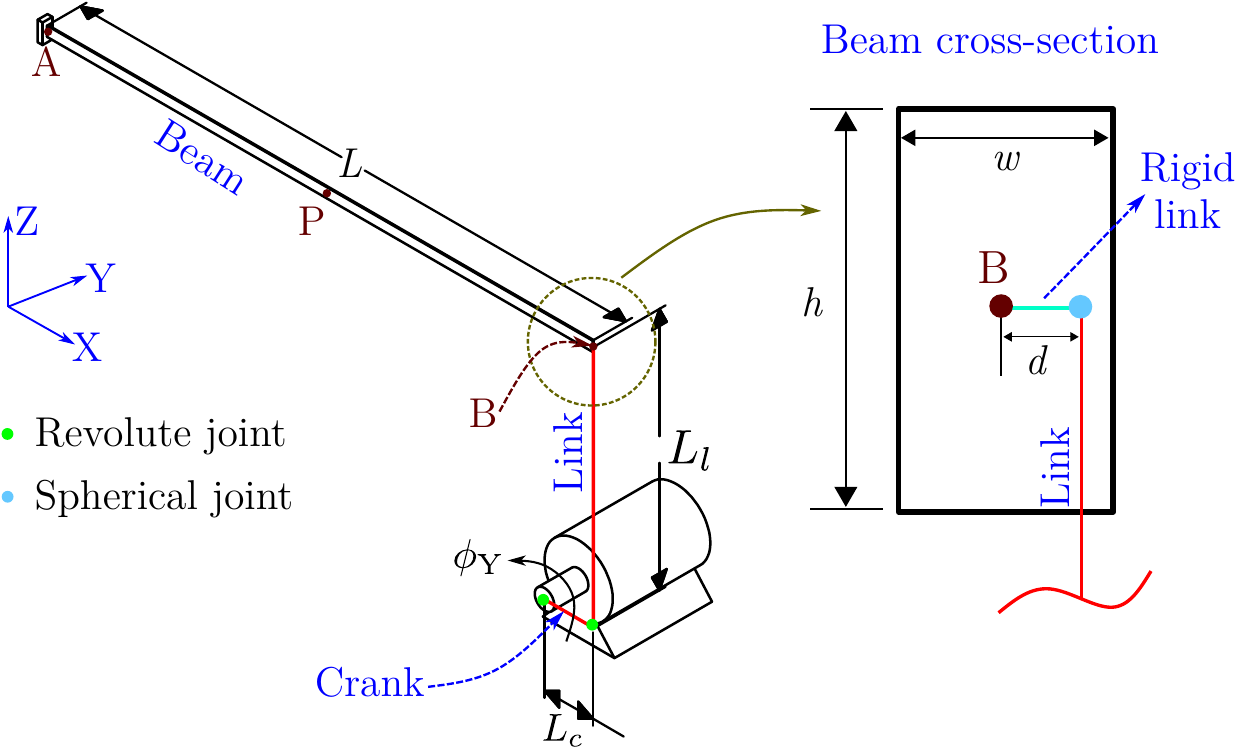}
\caption{The set-up of the lateral torsional buckling of a slender beam benchmark}
\label{fig:setupbuckle} 
\end{figure*}

\begin{figure*}
\centering
\includegraphics[width=1.0\textwidth]{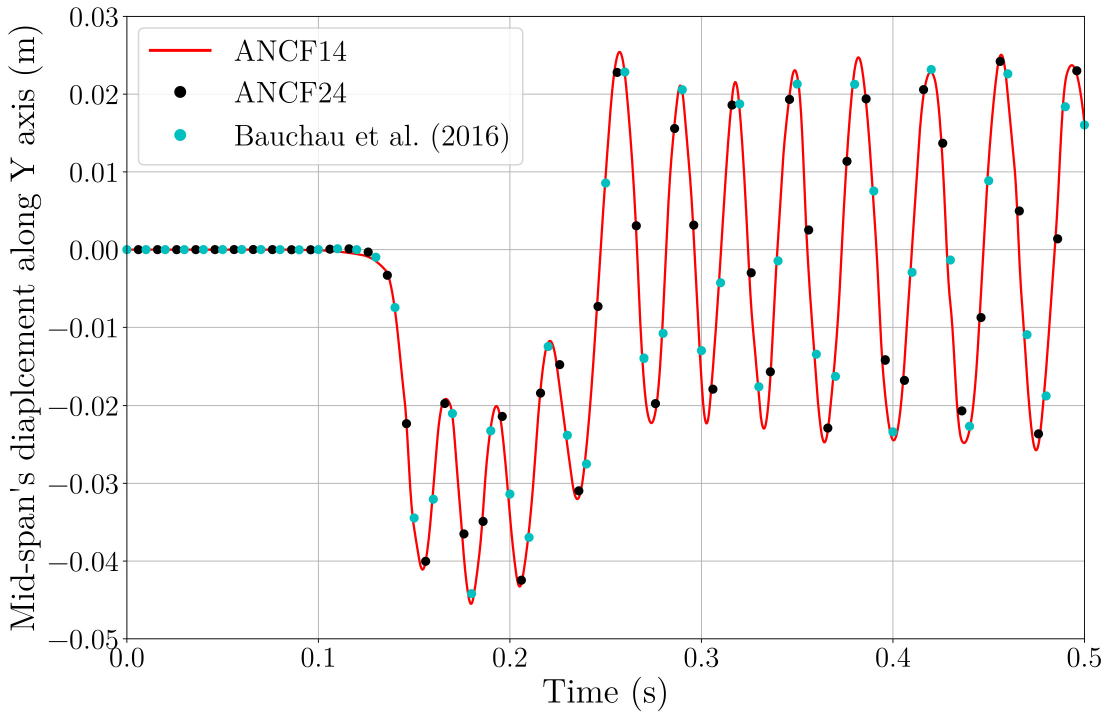}
\caption{\label{fig:u2dispbuckle} The displacement of Point P in time along the Y axis}
\end{figure*}

\begin{figure*}
\centering
\includegraphics[width=1.0\textwidth]{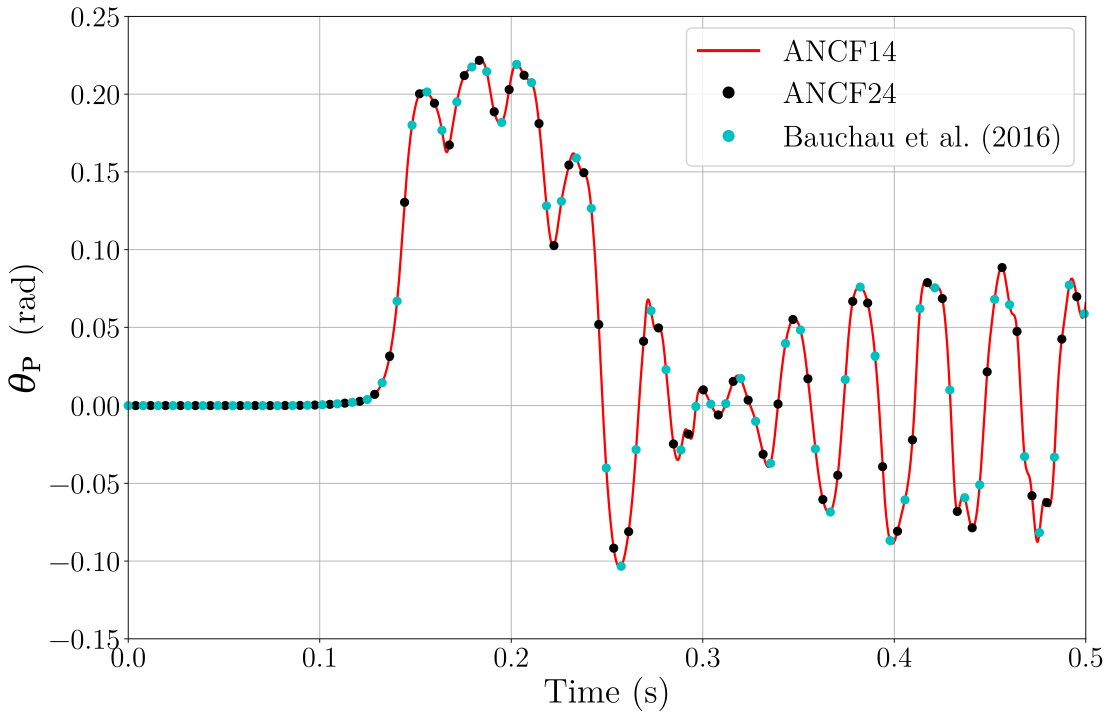}
\caption{\label{fig:twistbuckle} The cross-section's rotation $\theta_{\mathrm{P}}$ in time}
\end{figure*}

\begin{figure*}
\centering
\includegraphics[width=1.0\textwidth]{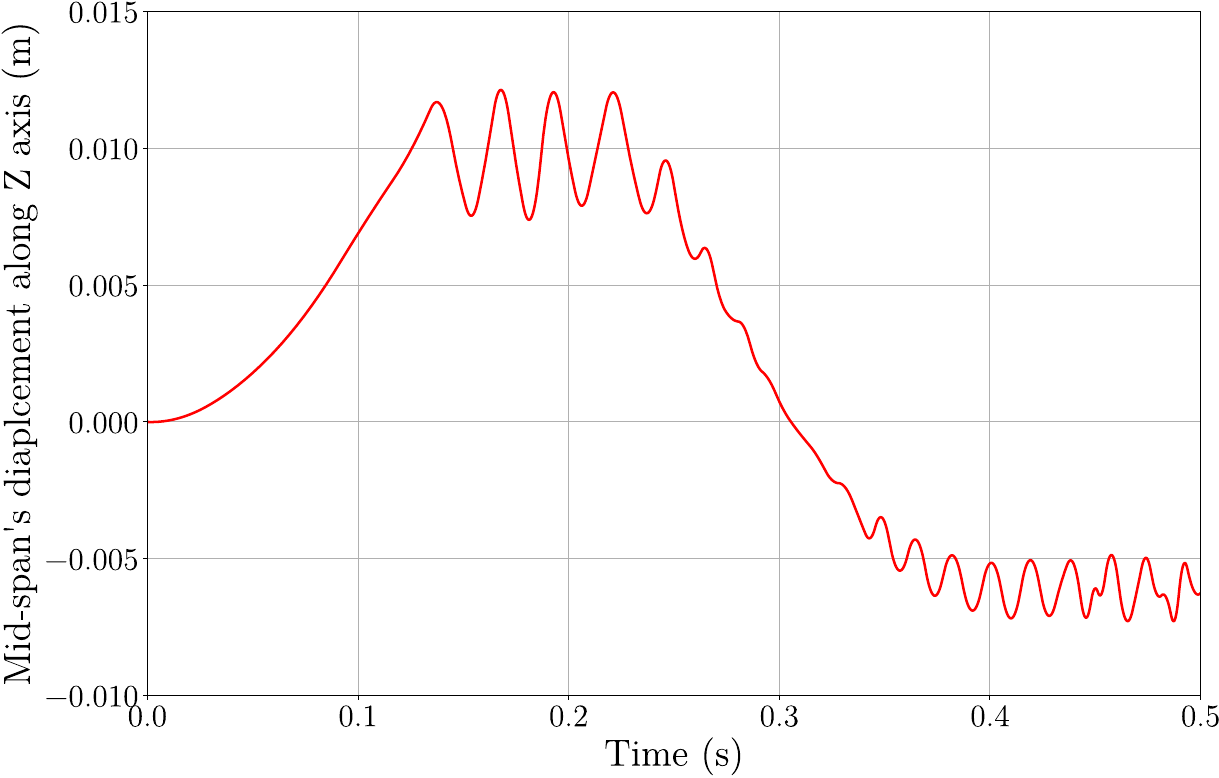}
\caption{\label{fig:u3dispbuckle} The displacement of Point P in time along the Z axis}
\end{figure*}

\begin{figure*}
\centering
\includegraphics[width=1.0\textwidth]{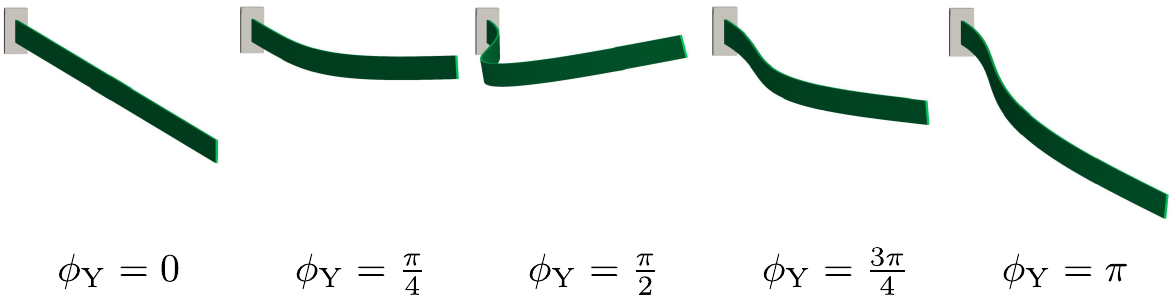}
\caption{\label{fig:snapshots} The beam at different $\phi_{\mathrm{Y}}$ in the lateral torsional buckling test }
\end{figure*}

When a beam is subject to a pure transverse load and bends, either its top or bottom face orthogonal to the load undergoes compression depending on the direction of the applied load. If the load is greater than a certain limit and the beam is \emph{unrestrained} laterally, the compressed face would buckle locally and due to the tension of the other face, the cross-section would twist about the beam's longitudinal axis, leading to both lateral and torsional deformations \cite{trahair2017flexural}. This phenomenon is called \emph{lateral torsional buckling} and must be accounted for in the design stage, as it considerably reduces the structure's capacity in supporting the applied loads. \\
\indent Considering its importance and the complex behavior of a beam under such circumstances, the last benchmark is dedicated to studying this event. Originally introduced in \cite{bauchau2016validation} and depicted in Figure \ref{fig:setupbuckle}, in this example, the tip of a slender beam is subject to an upward displacement applied through a link-crank mechanism. The beam has a length of $L$ and a uniform rectangular cross-section with a height $h$ and width $w$. It is clamped to the ground from End A and is connected to the link via a rigid connection and a spherical joint at End B. The rigid connection is off by a small distance $d$ from the beam's center-line in order to trigger the torsional buckling. The link and crank have uniform circular cross-sections with radii $r_{\mathrm{l}}$ and $r_{\mathrm{c}}$, respectively. Their lengths are $L_{\mathrm{l}}$ and $L_{\mathrm{c}}$. All components are made of aluminum with Young's modulus $E$, Poisson's ration $\nu$ and density $\rho$. The values of the parameters are provided in Table \ref{table:tablebuckling}. The crank is connected to the link and a rotor through revolute joints. The following rotation about the global Y axis is prescribed to the crank

\begin{equation}
\phi_{\mathrm{Y}}(t) = \begin{cases}
0.5 \pi \left(1- \cos \left(\pi t / T \right) \right) & \quad 0 \le t < T\\
\pi & \quad T \le t \le 0.5
\end{cases}
\end{equation}

\noindent where $T=0.4\:$s. Figure \ref{fig:u2dispbuckle} compares the plot of the beam's mid-span (Point P) displacement along the global Y axis and those reported in \cite{bauchau2016validation} and obtained by ANCF24 . Accordingly, the results are in excellent agreement. In terms of simulation time, the test with ANCF14 elements runs approximately 51 percent faster than the one with ANCF24 elements.  The evolution of the cross-section's rotation ($\theta_{\mathrm{P}}$) at Point P is illustrated in Figure \ref{fig:twistbuckle}. In this aspect too, ANCF14 can match the results of ANCF24 and \cite{bauchau2016validation}. \\
\indent Figure \ref{fig:u3dispbuckle} shows the displacement of Point P along the global Z axis in time. As can be observed, at the beginning of the process, the beam's mid-span displacement is only along the Z axis and the cross-section has no twist at that location. However, at about time $t \approx 0.12\:$s when the displacement threshold has reached, the beam buckles and starts simultaneously twisting and moving laterally along Y and Z axes. Figure \ref{fig:snapshots} depicts the beam at different $\phi_{\mathrm{Y}}$. Note that the test is run assuming large deformations.

\begin{figure*}
\centering
	\begin{subfigure}{1.0\textwidth}
        \includegraphics[width=\textwidth]{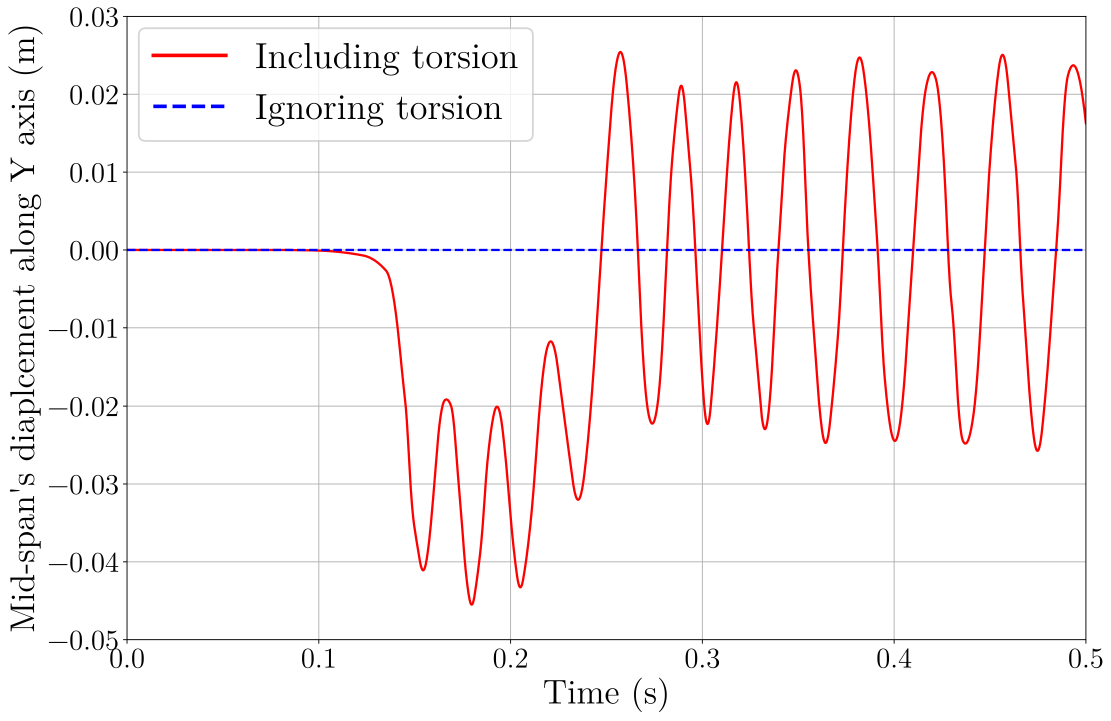}
        \caption{}
        \label{subfig:u2notwistBuckle}
    \end{subfigure}
    \begin{subfigure}{1.0\textwidth}
        \includegraphics[width=\textwidth]{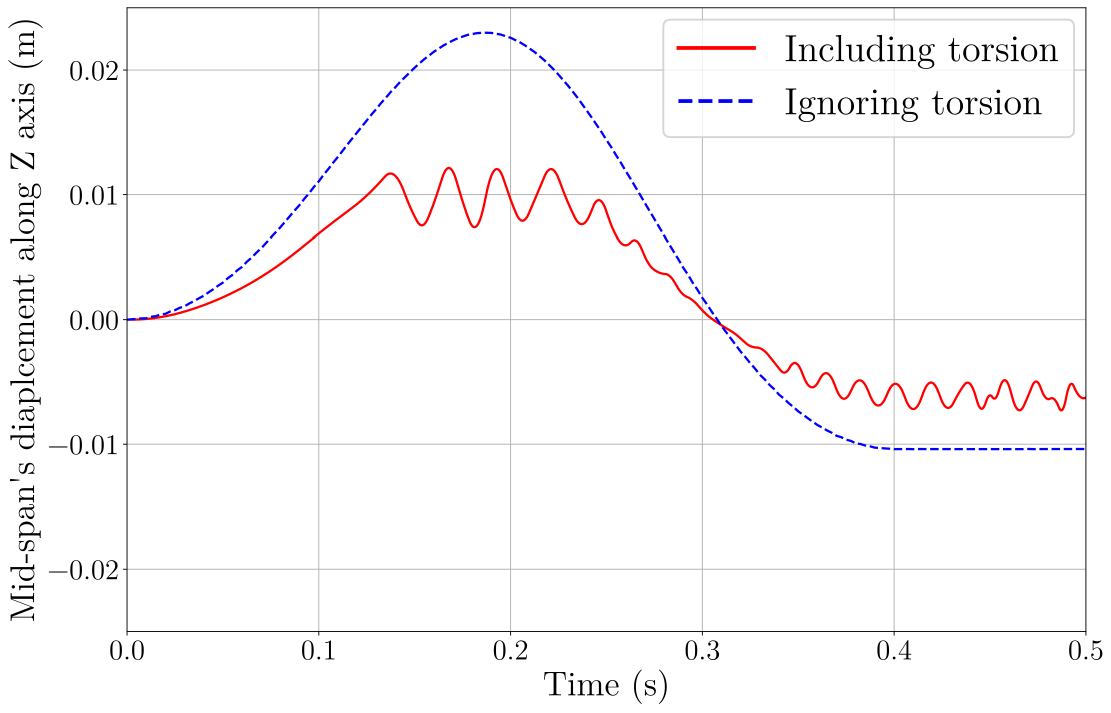}
        \caption{}
        \label{subfig:u3notwistBuckle}
    \end{subfigure}
    \caption{The displacement of Point P with and without including torsional effects}
    \label{fig:u2u3notwist} 
\end{figure*}

To show the significance of including torsion in the overall behavior of the beam, this problem is run disregarding the torsional effects. Figure \ref{fig:u2u3notwist} plots the displacement of Point P with this assumption against the results obtained previously. Note that the behaviors are substantially different. Since the torsional deformations are ignored, the beam does not buckle, so it has no displacement along the Y axis throughout the simulation. The displacement profile along the Z axis suggests a stable behavior with no oscillations and excessive movements throughout the entire simulation.

\FloatBarrier
\section{Conclusion}
\label{sec:conclusion}

In this paper, we have proposed a new low-order Euler-Bernoulli beam element capable of handling torsional deformations leveraging the absolute nodal coordinate formulation and the Bishop frame. This new element is called ANCF14 and has two nodes, each with 7 DOF that are global nodal coordinates and slopes as well as their cross-sectional rotation about the center-line. To describe its elastic deformations, the Bishop instead of the Serret-Frenet frame is chosen. It is shown that the former, unlike the latter, is always well-defined along the beam, thus leading to a singularity-free model even in the zero-curvature segments of the beam. Using the Bishop frame, the equations for ANCF14's elastic energy and its virtual work for small and large deformations are presented. 

To assay the newly developed element, four numerical examples are provided. In all these examples, the beams under study are subject to a combination of longitudinal, transverse bending and torsional deformations, and failing to capture them accurately would completely change their ultimate simulated behavior. It is shown that the results of employing ANCF14 are in excellent agreement with the theoretical and numerical results of other beam formulations reported in similar studies. From the computational efficiency standpoint, it is shown that for all the provided numerical examples the simulation times using ANCF14 are considerably (about 50 percent) lower than those with ANCF24 elements. This difference is due to the fewer degrees of freedom of ANCF14 compared to ANCF24 elements which is of more computational significance for larger problems.

For the sake of brevity, in this paper, the beam is considered straight in its undeformed configuration. This is not, however, a limiting assumption and the proposed element can be easily extended to handle initially curved beams with some minor modifications, namely considering $x$ in Figure \ref{fig:ANCF14} as the arc length from Node $i$ rather than the Euclidean distance from it. Also, ANCF14 is considered two-noded. It is straightforward to create a three-noded version of it. This would enable the quadratic interpolation of the cross-sectional rotation which may provide a more accurate analysis of torsional deformations.

The application of beam-like components in engineering structures is ubiquitous. The main intent of modeling these three-dimensional parts by beam finite elements is to somewhat alleviate the computational burden of simulating large scale real-world problems. This, however, comes with a compromise in the quality of their numerical solutions. Thus, in developing new element models, it is crucial to take into account both the computational efficiency and numerical accuracy at the same time. The current work demonstrates that ANCF14 can be beneficial in both regards.  

\newpage
\begin{appendices}

\section{}
\label{appen:ularge}

Putting Eq. (\ref{eq:finalgreenelems}) into Eq. (\ref{eq:totalelastic}) and defining $\bar{\varepsilon} := \| \vec{r}' \|^2 - 1$ we have

\begin{equation}\label{eq:elasticlarge}
\begin{aligned}
U_{\text{elastic}} &= \frac{1}{2} \int_V E \varepsilon_{11}^2 + 4 G \varepsilon_{12}^2 + 4 G \varepsilon_{13}^2 \: dV = \frac{1}{2} \int_V E \varepsilon_{11}^2 + G \tau_{\text{m}}^2 \left( \bar{y}^2 + \bar{z}^2 \right) \: dV \\
&= \frac{E}{8} \int_V \left[ \bar{\varepsilon} + \left( \gamma_1 \bar{y} + \gamma_2 \bar{z} \right)^2 + 2 \| \vec{r}' \| \left( \gamma_1 \bar{y} + \gamma_2 \bar{z} \right) + \tau_{\text{m}}^2 \left( \bar{y}^2 + \bar{z}^2 \right) \right]^2 \: dV \\ 
&+ \frac{G}{2} \int_V \tau_{\text{m}}^2 \left( \bar{y}^2 + \bar{z}^2 \right) \: dV \\
&= \frac{E}{8} \int_V \left[ \bar{\varepsilon}^2 + \left( \gamma_1 \bar{y} + \gamma_2 \bar{z} \right)^4 + 4 \| \vec{r}' \|^2 \left( \gamma_1 \bar{y} + \gamma_2 \bar{z} \right)^2 + \tau_{\text{m}}^4 \left( \bar{y}^2 + \bar{z}^2 \right)^2 \right] \: dV \\
&+ \frac{E}{4} \int_V \left[ \bar{\varepsilon} \left( \gamma_1 \bar{y} + \gamma_2 \bar{z} \right)^2 + 2 \bar{\varepsilon} \| \vec{r}' \| \left( \gamma_1 \bar{y} + \gamma_2 \bar{z} \right) + \bar{\varepsilon} \tau_{\text{m}}^2 \left( \bar{y}^2 + \bar{z}^2 \right)       \right] \: dV \\
&+ \frac{E}{4} \int_V \left[ 2 \| \vec{r}' \|  \left( \gamma_1 \bar{y} + \gamma_2 \bar{z} \right)^3 + \tau_{\text{m}}^2 \left( \bar{y}^2 + \bar{z}^2 \right)  \left( \gamma_1 \bar{y} + \gamma_2 \bar{z} \right)^2       \right] \: dV \\
&+ \frac{E}{4} \int_V \left[  2 \| \vec{r}' \| \tau_{\text{m}}^2 \left( \bar{y}^2 + \bar{z}^2 \right) \left( \gamma_1 \bar{y} + \gamma_2 \bar{z} \right)   \right] \: dV + \frac{G}{2} \int_V \tau_{\text{m}}^2 \left( \bar{y}^2 + \bar{z}^2 \right) \: dV \\
\end{aligned}
\end{equation}

\noindent Provided the same assumptions as those considered for deriving Eq. (\ref{eq:totalelasticsmalldef}) and assuming the fourth moments of area

\begin{equation}
\int\limits_A \bar{y}^4 \: dA \approx 0, \: \int\limits_A \bar{z}^4 \: dA \approx 0, \int\limits_A \bar{y}^2 \bar{z}^2 \: dA \approx 0
\end{equation}

\noindent Eq. (\ref{eq:elasticlarge}) reduces to

\begin{equation}
\begin{aligned}
U_{\text{elastic}} &= \frac{E}{8} \int_V \bar{\varepsilon}^2 \: dV + \frac{G}{2} \int_V \tau_{\text{m}}^2 \left( \bar{y}^2 + \bar{z}^2 \right) \: dV \\
&+ \frac{E}{4} \int_V \left[ \bar{y}^2 \gamma_1^2 \left( 2\| \vec{r}' \|^2  + \bar{\varepsilon} \right) +  \bar{z}^2 \gamma_2^2 \left( 2\| \vec{r}' \|^2  + \bar{\varepsilon} \right) + \bar{\varepsilon} \tau_{\text{m}}^2 \left( \bar{y}^2 + \bar{z}^2 \right)			\right] \: dV \\
&= \frac{E}{8} \int_V \left(\| \vec{r} ' \|^2 -1\right)^2  \: dV \\
&+ \frac{E}{4} \int_V \bar{y}^2 \gamma_1^2 \left( 3 \| \vec{r}' \|^2 - 1 \right) \: dV + \frac{E}{4} \int_V \bar{z}^2 \gamma_2^2 \left( 3 \| \vec{r}' \|^2 - 1 \right) \: dV \\
&+ \frac{E}{4} \int_V \left( \bar{y}^2 + \bar{z}^2 \right) \tau_m^2 \left( \| \vec{r}' \|^2- 1 \right) \: dV + \frac{G}{2} \int_V \left( \bar{y}^2 + \bar{z}^2 \right) \tau_m^2  \: dV \\
&= \frac{EA}{8} \int_0^l \left(\| \vec{r} ' \|^2 -1\right)^2  \: dx \\
&+ \frac{EI_z}{4} \int_0^l \gamma_1^2 \left( 3 \| \vec{r}' \|^2 - 1 \right) \: dx + \frac{EI_y}{4} \int_0^l \gamma_2^2 \left( 3 \| \vec{r}' \|^2 - 1 \right) \: dx \\
&+ \frac{EJ_t}{4} \int_0^l \tau_m^2 \left( \| \vec{r}' \|^2- 1 \right) + \frac{GJ_t}{2} \int_0^l \tau_m^2  \: dx \: dx
\end{aligned}
\end{equation}

\section{}
\label{appen:derivative}

Based on Eq. (\ref{eq:position}), one has $\vec{r} = \mathbf{S} \vec{q}$ and consequently

\begin{equation}\label{eq:drprimedq}
\begin{aligned}
\frac{\partial \vec{r}'}{\partial \vec{q}} = \mathbf{S}' \\
\end{aligned}
\end{equation}

\noindent Also using Eq. (\ref{eq:torsiondef}) and Eq. (\ref{eq:rotationinterpolation}), $\partial \tau_m/\partial \vec{q}$ can be analytically calculated everywhere along the element through 

\begin{equation}\label{eq:dtaudq}
\begin{aligned}
\frac{\partial \tau_m}{\partial \vec{q}} = \frac{\partial \theta'}{\partial \vec{q}} = \mathbf{\bar{S}}'
\end{aligned}
\end{equation}

\noindent According to Eq. (\ref{eq:materialdarboux}), $\partial \gamma_1/\partial \vec{q}$ and $\partial \gamma_2/\partial \vec{q}$ are

\begin{equation}
\begin{aligned}
\gamma_1 = \vec{t}' \cdot \vec{y} & \implies  \frac{\partial \gamma_1}{\partial \vec{q}}  = \vec{y}^{\mathrm{T}}\frac{\partial \vec{t}'}{\partial \vec{q}}+ \vec{t}'^{\mathrm{T}} \frac{\partial \vec{y}}{\partial \vec{q}} \\
\gamma_2  = \vec{t}'\cdot \vec{z} & \implies  \frac{\partial \gamma_2}{\partial \vec{q}} = \vec{z}^{\mathrm{T}}\frac{\partial \vec{t}'}{\partial \vec{q}}+ \vec{t}'^{\mathrm{T}}\frac{\partial \vec{z}}{\partial \vec{q}}
\end{aligned}
\end{equation}

\noindent Knowing $\vec{t} = \vec{r}' / \| \vec{r}' \|$, for any point on the center-line, $\vec{t}'$ and $\partial \vec{t}' / \partial \vec{q}$ can be analytically computed via

\begin{equation}
\begin{aligned}\label{eq:dtdq}
\frac{\partial \vec{t}}{\partial \vec{q}} &= \frac{1}{\| \vec{r}' \|} \left( \mathbf{I} - \frac{\vec{r}' \vec{r}'^{\text{T}}}{\| \vec{r}' \|^2} \right) \frac{\partial \vec{r}'}{\partial \vec{q}} \\
\frac{\partial \vec{t}'}{\partial \vec{q}} &= \frac{1}{\| \vec{r}' \|} \left( \mathbf{I} - \frac{\vec{r}' \vec{r}'^{\text{T}}}{\| \vec{r}' \|^2} \right) \left( \frac{\partial \vec{r}''}{\partial \vec{q}}  - \frac{\vec{r}'^{\text{T}} \vec{r}''}{\| \vec{r}' \|^2} \frac{\partial \vec{r}'}{\partial \vec{q}}  \right) \\
&+ \frac{1}{\| \vec{r}' \|^3} \left( \frac{\vec{r}'^{\text{T}} \vec{r}''}{\| \vec{r}' \|^3} \vec{r}' \vec{r}'^{\text{T}} - \vec{r}'' \vec{r}'^{\text{T}} - \vec{r}' \vec{r}''^{\text{T}} \right) \frac{\partial \vec{r}'}{\partial \vec{q}} 
\end{aligned}
\end{equation}

\noindent where $\vec{r}' = \mathbf{S}' \vec{q}$ and $\vec{r}'' = \mathbf{S}'' \vec{q}$. Using Eq. (\ref{eq:materialfrombishop}), $\partial \vec{y} / \partial \vec{q}$  and $\partial \vec{z} / \partial \vec{q}$ become

\begin{equation}
\begin{aligned}
\frac{\partial \vec{y}}{\partial \vec{q}} &= -\frac{\partial \theta}{\partial \vec{q}} \sin (\theta) \vec{u} + \frac{\partial \theta}{\partial \vec{q}} \cos (\theta) \vec{v} + \cos (\theta) \frac{\partial \vec{u}}{\partial \vec{q}} + \sin (\theta) \frac{\partial \vec{v}}{\partial \vec{q}} \\
\frac{\partial \vec{z}}{\partial \vec{q}} &= -\frac{\partial \theta}{\partial \vec{q}} \cos (\theta) \vec{u} - \frac{\partial \theta}{\partial \vec{q}} \sin (\theta) \vec{v} - \sin (\theta) \frac{\partial \vec{u}}{\partial \vec{q}} + \cos (\theta) \frac{\partial \vec{v}}{\partial \vec{q}}
\end{aligned}
\end{equation}

\noindent where $\vec{u}$, $\vec{v}$ are obtained employing Algorithm \ref{alg:rotationBP}. Their derivatives $\partial \vec{u} / \partial \vec{q}$ and $\partial \vec{v} / \partial \vec{q}$, therefore, can be resulted by the algorithm below.

\vspace{1em}
\begin{algorithm}[H]
\caption{Calculating $\partial \vec{u} / \partial \vec{q}$ and $\partial \vec{v} / \partial \vec{q}$ of the Bishop frame for a selected number of points along a spatial curve}
\SetAlgoLined
\KwInput{Curve equation $\vec{r}(x)$, $\left\{ \vec{t}^{0}, \vec{u}^{0}, \vec{v}^{0} \right\}$ and the desired $x_i$ locations}
\KwOutput{$\partial \vec{u}^{i} / \partial \vec{q}, \partial \vec{v}^{i} / \vec{q}$ for $i=1,\cdots,N$}
 \For{$i=1, \cdots, N$}{
  $\vec{t}^{i} \leftarrow \vec{r}^{'}(x_i) / \| \vec{r}^{'}(x_i)\|$\;
  $\vec{n} \leftarrow \vec{t}^{i-1} \times \vec{t}^{i}$\;
  \eIf{$\| \vec{n} \| = 0$}{
   $\partial \vec{u}^{i} / \partial \vec{q} \leftarrow \partial \vec{u}^{i-1} / \partial \vec{q}$\;
   $\partial  \vec{v}^{i} / \partial \vec{q} \leftarrow \partial  \vec{v}^{i-1} / \partial \vec{q}$\;
   }{
   $\hat{\vec{n}} \leftarrow \vec{n} / \| \vec{n} \|$\;
   Compute $\partial \vec{t}^i / \partial \vec{q}$ from Eq. (\ref{eq:dtdq}) \;
   $\partial \vec{n} / \partial \vec{q} \leftarrow \partial \vec{t}^{i-1} / \partial \vec{q} \times \vec{t}^{i} + \vec{t}^{i-1} \times \partial \vec{t}^{i} / \partial \vec{q}$\;
   $\partial \hat{\vec{n}}  / \partial \vec{q} \leftarrow \left( 1 / \| \vec{n} \| \right) \left( \mathbf{I} - \vec{n} \vec{n}^{\text{T}} / \| \vec{n} \|^2 \right) \left( \partial \vec{n} / \partial \vec{q} \right)$\;
   $\phi \leftarrow \arccos \left(\vec{t}^{i-1} \cdot \vec{t}^{i} \right)$\;
   $\partial \phi / \partial \vec{q} \leftarrow -\left( \vec{t}^{i^{\text{T}}} \left( \partial \vec{t}^{i-1} / \partial \vec{q} \right) + \vec{t}^{{i-1}^{\text{T}}} \left( \partial \vec{t}^{i} / \partial \vec{q} \right)  \right) / \sqrt{1 - \vec{t}^{i-1} \cdot \vec{t}^{i}}$\;
   $\vec{u}^i \leftarrow \vec{u}^{i-1} \cos(\phi) + \left(\hat{\vec{n}} \times \vec{u}^{i-1} \right) \sin(\phi) + \hat{\vec{n}} \left( \hat{\vec{n}} \cdot \vec{u}^{i-1} \right) \left( 1- \cos(\phi) \right) $\;
   $\vec{v}^i \leftarrow \vec{v}^{i-1} \cos(\phi) + \left(\hat{\vec{n}} \times \vec{v}^{i-1} \right) \sin(\phi) + \hat{\vec{n}} \left( \hat{\vec{n}} \cdot \vec{v}^{i-1} \right) \left( 1- \cos(\phi) \right) $\;   
   Compute $\partial \vec{u}^i / \partial \vec{q}$ and $\partial \vec{v}^i / \partial \vec{q}$ accordingly\;
  }
 }
\end{algorithm}

\vspace{1em}

\noindent where $\partial \vec{t}^{i-1} / \partial \vec{q} \times \vec{t}^{i}$ means the cross-product of each column of  $\partial \vec{t}^{i-1} / \partial \vec{q}$ by $\vec{t}^{i}$ from right; and $\vec{t}^{i-1} \times \partial \vec{t}^{i} / \partial \vec{q}$ is the cross-product of $\vec{t}^{i-1} $ by each column of $\partial \vec{t}^{i} / \partial \vec{q}$ from left.

\newpage

\end{appendices}

\newpage
\bibliography{main}
\bibliographystyle{spmpsci}

\end{document}